\documentclass[apj]{emulateapj}
\slugcomment{{\sc Accepted for publication in ApJ}} 

\pdfoutput=1 

\usepackage{graphicx}
\usepackage{epstopdf}
\usepackage{multirow}
\usepackage{lscape}

\slugcomment{Accepted for publication in ApJ}

\shorttitle{The KONA survey}
\shortauthors{M\"uller S\'anchez et al.}

\begin{document}


\title{The Keck/OSIRIS Nearby AGN Survey (KONA) I. The Nuclear $K-$band Properties of Nearby AGN$^1$
}


\author{F. M\"uller-S\'anchez$^{1}$, E. K. S. Hicks$^{2}$, M. Malkan$^{3}$, R. Davies$^{4}$, P. C. Yu$^{3, 5}$, S. Shaver$^{1}$, B. Davis$^{1, 6}$}

\affil{$^1$ Department of Astrophysical and Planetary Sciences, University of Colorado, Boulder, CO 80309, USA}

\affil{$^2$ Department of Physics and Astronomy, University of Alaska Anchorage, AK 99508, USA}

\affil{$^3$ Department of Physics and Astronomy, University of California, Los Angeles, CA 90095-1562, USA}

\affil{$^4$ Max-Planck-Institut f\"ur Extraterrestrische Physik, Postfach 1312, D-85741 Garching, Germany}

\affil{$^5$ Graduate Institute of Astronomy, National Central University, 300 Jhongda Road, Jhongli 32001, Taiwan}

\affil{$^6$ Department of Astronomy and Astrophysics, Pennsylvania State University, PA 16802, USA}



\footnotetext[1]{Based on observations at the W. M. Keck Observatory, which is operated as a scientific partnership among the California Institute of Technology, the University of California, and the National Aeronautics and Space Administration. The observatory was made possible by the generous financial support of the W. M. Keck Foundation.}






\begin{abstract} 

We introduce the Keck Osiris Nearby AGN survey (KONA), 
a new adaptive optics-assisted integral-field spectroscopic survey of Seyfert galaxies. 
KONA permits at $\sim0.1\arcsec$ resolution a detailed study 
of the nuclear kinematic structure of gas and stars in a representative sample of 40 local bona fide active galactic nucleus (AGN). 
KONA seeks to characterize 
the physical processes responsible for 
the coevolution of supermassive black holes and galaxies, principally inflows and outflows. With these IFU data of the nuclear regions of 40 Seyfert galaxies, the KONA survey will be able to study, for the first time, a number of key topics with meaningful statistics. In this paper we study the nuclear $K-$band properties of nearby AGN. We find that the $K-$band ($2.1$ $\mu$m) luminosities of the compact Seyfert 1 nuclei are correlated with the hard X-ray luminosities, implying a non-stellar origin for the majority of the continuum emission. The best-fit correlation is log$L_K\,=0.9$log$L_{\mathrm{2-10\, keV}}\,+4$ over three orders of magnitude in both $K-$band and X-ray luminosities. 
We find no strong correlation between $2.1$ $\mu$m luminosity and hard X-ray luminosity for the Seyfert 2 galaxies. The spatial extent and spectral slope of the Seyfert 2 galaxies indicate the presence of nuclear star formation and attenuating material (gas and dust), which in some cases is compact and in some galaxies extended. We detect coronal-line emission in 36 galaxies and for the first time in five galaxies. Finally, we find 4/20 galaxies that are usually classified as Seyfert 2 based on their optical spectra exhibit a broad component of Br$\gamma$ emission, and one galaxy (NGC 7465) shows evidence of a double nucleus.

\end{abstract}




\section{Introduction}\label{introduction}

A fundamental question in astrophysics today is understanding the physical processes that govern
the coevolution of  supermassive black holes (SMBHs) and galaxies, 
resulting in surprisingly tight correlations such as the M$_{\mathrm{BH}}-\sigma^*$ relation (see Heckman \& Best 2014 for a review). Active galactic nuclei (AGNs) have emerged as likely players in this coevolution, since they are by definition actively accreting material in the centers of galaxies. Independently of the precise physical mechanism of inflow, mass accretion to the central region builds up both the galaxy stellar mass and the SMBH mass. AGN feedback is then commonly invoked to suppress star formation in a galaxy, shape the galaxy luminosity function and control the growth of the SMBH (e.g., Di Matteo et al. 2005, Somerville et al. 2008, Hopkins \& Elvis 2010). However, the observations are still unclear about whether this combination of central inflows, and AGN feedback regulating SMBH and bulge growth, is indeed the driver behind galaxy - SMBH coevolution. 

This is mainly because observations have focused on processes occurring at kpc-scales (the transport of gas from a few kpc to scales of a few hundreds parsecs) and on integrated galaxy properties, 
rather than on how the material actually flows toward the SMBH on the $<200$ pc spatial scales relevant to the time scales of the AGN lifetime and the onset of outflows (Schawinski et al. 2012, Comerford et al. 2017). 
Due to the small spatial scales involved, it is nearby AGN that offer the only opportunity to characterize the inflow and outflow processes in detail and test the methods used in numerical simulations and theoretical models. 
Nearly all of the nearby AGN up to distances of $\sim150$ Mpc are classified as Seyfert galaxies, which are believed to be the counterparts of powerful quasi-stellar objects (QSOs) and quasars in the local universe, following the same unification scheme and exhibiting the same physical processes 
(Antonucci 1993, Krolik 1998, Elitzur \& Shlossman 2006, Hopkins et al. 2006). 
Indeed recent simulations and observations suggest that, even at $z = 2$, disk processes, characteristic of the majority of Seyfert galaxies, provide considerable AGN fueling (Schawinski et al. 2012, Villforth et al. 2014), and AGN-driven outflows have been observed in local Seyfert galaxies and galaxies at high-redshifts (M\"uller-S\'anchez et al. 2011, Cresci et al. 2015). 

 
In order to characterize the spatial distribution and kinematics of the molecular and ionized gas within the central 200 pc of AGN, it is necessary to have high signal-to-noise (S/N) spectra with high spatial resolution over a two-dimensional field of view. 
This is possible in the near-IR by combining the power of adaptive optics (AO) with integral-field spectroscopy (IFS) in 8 m class telescopes. The $K-$band
is best suited to accomplish this, since the $2.12$ $\mu$m molecular hydrogen emission (a tracer of the dense interstellar medium [ISM]), $2.16$ $\mu$m Br$\gamma$ (a tracer of the low-ionization gas and the narrow line region), $1.97$ $\mu$m [Si VI] (a tracer of the highly ionized gas and the coronal line region), and stellar 2.3 $\mu$m CO bandheads are observable. 

The first AO-assisted integral-field spectroscopic study of a Seyfert galaxy (the Circinus galaxy) was presented in M\"uller-S\'anchez et al. (2006). Since then, several studies of individual Seyfert galaxies and small samples of nearby AGNs with AO-assisted IFS have offered new understandings of AGN-galaxy coevolution, in particular (i) the presence of star formation in the central $<50$ parsecs of the galaxy and its influence on feeding the AGN, (ii) the inflow of gas toward the nuclear region, (iii) the properties of the molecular gas and its relation to the torus 
and (iv) the characteristics of the outflows of molecular/ionized gas and their interaction with
the ISM (M\"uller-S\'anchez et al. 2006, Riffel et al. 2006, Davies et al. 2007, Neumayer et al. 2007, Riffel et al. 2008, Riffel et al. 2009, M\"uller-S\'anchez et al. 2009, Davies et al. 2009, Hicks et al. 2009, Friedrich et al. 2010, Storchi-Bergmann et al. 2010, M\"uller-S\'anchez et al. 2011, Riffel \& Storchi-Bergmann 2011a, Riffel \& Storchi-Bergmann 2011b, Storchi-Bergmann et al. 2012, Iserlohe et al. 2013, Mazzalay et al. 2013, M\"uller-S\'anchez et al. 2013, Riffel et al. 2013, U et al. 2013, Greene et al. 2014, Fischer et al. 2015, Riffel et al. 2015, M\"uller-S\'anchez et al. 2016, Fischer et al. 2017, Rodr\'iguez-Ardila et al. 2017, Sch\"onell et al. 2017, Riffel et al. 2017, Diniz et al. 2017).

Despite the considerable success of AO-assisted IFS for the study of active galaxies, this observational technique has scarcely been used to explore medium/large samples of AGN.
This is mainly due to the large amount of telescope time needed to observe such objects as well as the intrinsic difficulties of conducting detailed systematic analyses with this technique. 
The only exceptions are the studies of Burtscher et al. (2015) and Davies et al. (2015). Burtscher et al. (2015) selected observations of all AGN in the European Southern Observatory (ESO) / SINFONI archive to study the properties of the spatially unresolved near-infrared non-stellar continuum. However, this study does not address spatially resolved structures or kinematics, some of the galaxies were not observed with AO, and the sample of AGN is very heterogeneous, containing Seyfert galaxies, low-ionization nuclear emission-line region (LINER) galaxies, low-luminosity AGN, and even ultraluminous infrared galaxies (ULIRGs). Davies et al. (2015) presents a complete volume limited sample of 20 $SWIFT$ Seyfert galaxies and a complementary matched sample of inactive galaxies observed with AO and the instrument SINFONI on the Very Large Telescope (VLT). The scientific goals of this project consist primarily of finding the differences between active and inactive galaxies and the triggering mechanisms of nuclear activity (see also Sosa-Brito et al. 2001, Hicks et al. 2013, Davies et al. 2014 and Rothberg et al. 2015 for examples of IFU surveys of AGN but without AO).

In this paper we present the Keck/OSIRIS Nearby AGN (KONA) survey, which will investigate the nuclear spatial distribution and kinematics of gas and stars in an unprecedented sample of 40 bona fide nearby AGN observed with AO-assisted IFS. 
KONA not only provides the means to statistically address topics related to inflow and outflow characteristics of Seyfert galaxies (rather than being biased by small
number statistics and the significant variations from one galaxy to another), but also makes possible a robust investigation of potential trends with AGN power, Eddington ratio, Seyfert type, and other properties of the AGN and its host galaxy.
In Section 2 we summarize the key scientific goals of the KONA survey.  
We then present a detailed description of the sample selection and sample characteristics in Section 3. A brief overview of the KONA instrumentation and data reduction is given in Section 4. 
We demonstrate KONA's scientific potential by showing nuclear spectra of all KONA galaxies and the measured fluxes of emission lines in Section 5. Measurements of the AGN continuum in the $K-$band are also presented in this section. In Section 6 we compare the $K-$band luminosity of the KONA galaxies with their hard X-ray luminosity and find a good correlation between these two quantities (particularly for Seyfert 1s). A summary of the survey and the results is presented in Section 7.

Throughout this paper, we use a flat $\Lambda$CDM cosmology with $\Omega_m=0.27$, $\Omega_\Lambda=0.73$, and $H_0= 70$ km s$^{-1}$ Mpc$^{-1}$.

\section{Scientific Goals}\label{goals}

KONA seeks to do a full characterization of the physical processes occurring in the central 200 pc of nearby AGN. 
KONA's power resides in the possibility of doing this in a moderately large sample of AGN, probing a wide range of environment, AGN luminosity, and Eddington ratio (see Section 3). 
KONA's sample size of 40 AGN, combined with its ability to map the 2D properties of the gas and the stars at nearly the diffraction limit of an 8 m class telescope, promises revolutionary discoveries in the field of AGN research.

The primary scientific goals of this program are to provide galaxy formation and evolution models with
observational constraints on (1) the processes responsible for driving gas inward to the SMBH, (2) the processes
driving the observed outflows and the potential for meaningful feedback, (3) the role of the nuclear molecular gas in obscuring the AGN, and (4) a comparison of nuclear properties in subsamples (e.g., Seyfert 1s versus 2s) and across the range of AGN luminosities (i.e., dependencies of inflow/outflow rates on the AGN luminosity or Eddington ratio). These are discussed in
more detail below.

\begin{figure*}
\epsscale{.99}
\plotone{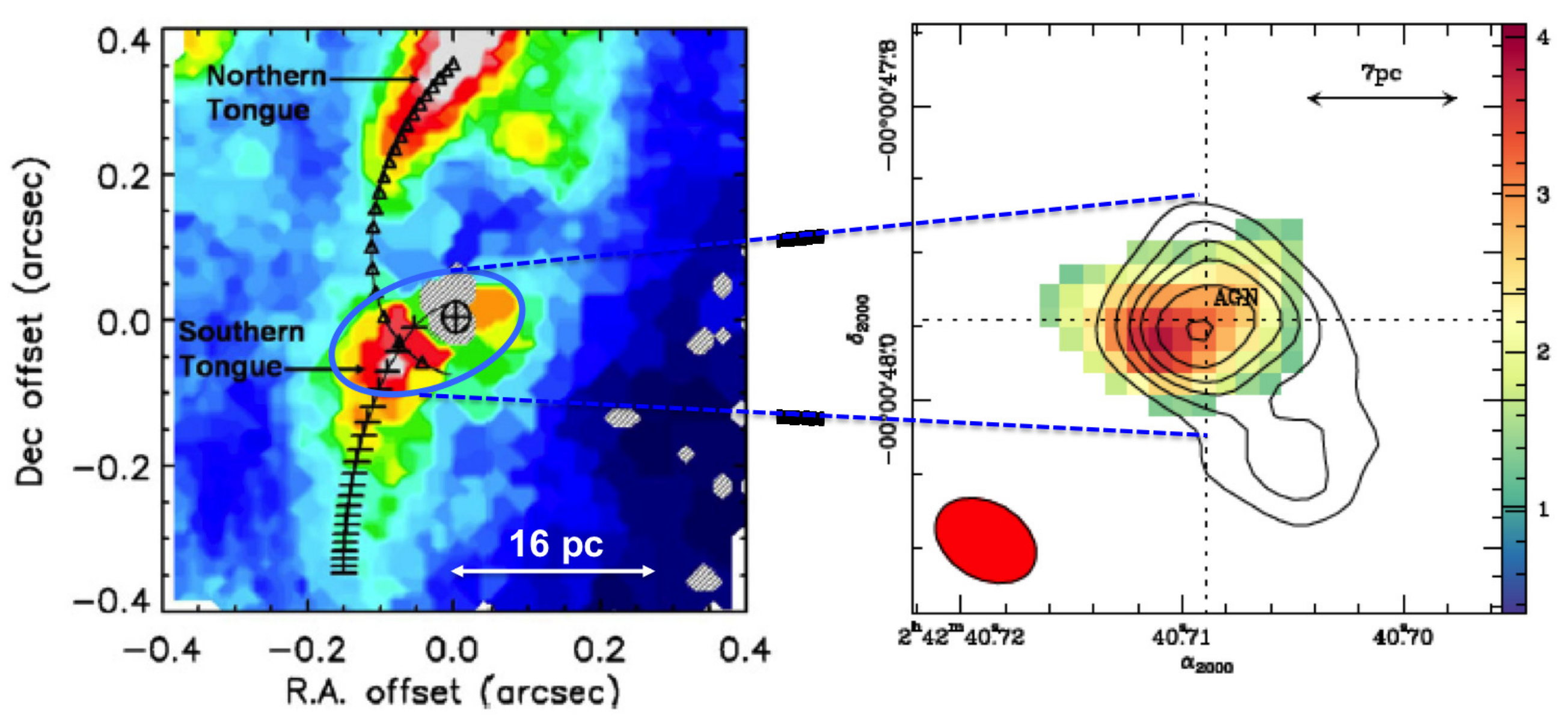}
\caption{{\it Left:} SINFONI flux map of H$_2$ emission in the central 50 pc of NGC1068 (from
M\"uller-S\'anchez et al. 2009). The central structure is interpreted as the outer part of the torus. Radially inflowing clouds are present in the torus. The open triangles show the projected trajectory of the northern streamer of gas. The half-crosses show the past trajectory of the gas which is currently located in front of the AGN. {\it Right:} ALMA image of CO emission in the nuclear region of NGC 1068 (Garcia-Burillo et al. 2016). The contours show emission at 250 GHz (cold dust).
\label{fig1}}
\end{figure*}

\subsection{Inflows - What Drives Gas from Hundred-parsec Scales into the Nucleus?}\label{inflows}

Several mechanisms for bringing gas into the nuclear regions
of galaxies have been suggested; however, there is no single mechanism for the entirety of the galaxy. 
While mergers and bars can generate gas reservoirs and a starburst region in the central kiloparsec, 
the primary physical mechanism for bringing gas
to the environs of the central SMBH has not been identified yet.

The inward flow of molecular gas on 10s of parsec scales has been revealed in a number of galaxies by
means of near-IR IFUs, either as inflowing material along a nuclear spiral (NGC 4051: Riffel et al. 2008, NGC 1097: Davies et al. 2009, Mrk 1066: Riffel et al. 2012, and Mrk 79: Riffel et al. 2013), a nuclear bar (NGC 4388: Greene et al. 2014, NGC 3081: M\"uller-S\'anchez et al. 2014; see also Schnorr-M\"uller et al. 2016), or as
streaming motions in highly eccentric orbits (NGC 1068: M\"uller-S\'anchez et al. 2009). In Davies et al. (2014), we found that NGC 3227 and NGC 5643 also support the presence of molecular bars in the
nuclear regions of Seyfert galaxies. In addition, this study provided evidence of
at least two galaxies in which counter-rotating disks of gas suggest stochastic inflow into the nuclear region.  
Although better statistics are needed to determine the significance of each of these fueling mechanisms, these results suggest that disk processes on $< 200$ pc scales are crucial in driving gas inward. However, their connection to AGN accretion is not well understood yet. 
Probing the gas kinematics directly in different types of galactic nuclei (type 1 vs type 2 and a wide range of luminosities)
will reveal how ubiquitous gas streaming really is, as well as whether the gas inflow rates (obtained from
modeling of the kinematics) are related to the accretion or star formation rates (SFRs). The dependence of the mechanism driving inflow with the host galaxy type and environment will be also investigated (Hicks et al. 2018, in preparation). 

\subsection{Outflows - How Do Accreting Black Holes Influence Their Host Galaxies?}\label{outflows}

M\"uller-S\'anchez et al. (2011) found that AGN-driven outflows of ionized gas are prevalent throughout nearby AGN, and that biconical models of radial outflow provide a good fit to the spatially resolved kinematics. The mass outflow rates are $2-3$ orders of magnitude greater than the SMBH accretion rates, but are comparable to the estimated inflow rates to the central 25 pc, suggesting that AGN feedback prevents gas from reaching the SMBH. 
Furthermore, the kinetic luminosity of the outflow follows a relationship with the velocity dispersion of the molecular gas in the central 25 pc, but this becomes indiscernible when the large-scale dispersion is considered. This suggests that AGN feedback has a strong impact on the turbulent ISM near the SMBH (probably disrupting the conditions necessary for star formation), but not on the galaxy disk. With this statistically powerful meaningful sample of nearby AGN, we will be able to make a definitive assessment of the
impact of the interaction of AGN-driven outflows with the ISM (negative or positive feedback), as well as
compare the outflow mass to the AGN luminosity and accretion rate, to ascertain under what
circumstances the AGN can drive a significant outflow that modifies the evolution of the host galaxy in a
meaningful way.

\subsection{The Molecular Torus and Unification Schemes}\label{torus}

The most popular version of the unified model of AGN invokes the existence of a geometrically and optically thick torus, made from dense molecular gas and dust, which causes the aspect-angle obscuration of AGNs.
Due to the small spatial scales involved, direct imaging is challenging. Therefore, most of its properties are poorly understood. 
The torus in NGC 1068 is arguably the only torus in the universe with a direct image. M\"uller-S\'anchez et al. (2009) interpreted the 2D properties of the warm molecular gas at $2.12$ $\mu$m as the outer part of the torus (with a size of $\sim10$ pc). Its kinematics exhibits two components: radial inflow and rotation (Figure 1). The same structure (size, orientation, offset peak of molecular emission, and kinematics) has been recently observed with ALMA (Garcia-Burillo et al. 2016), confirming that the warm molecular gas (H$_2$) can be a good tracer of the CO torus. The molecular gas structure revealed by SINFONI and ALMA has the same orientation as the H$_2$O masers, the nuclear emission at 5 GHz and the mid-IR Gaussian disk detected with MIDI at sub-pc scales (Figure 1). 
This suggests that the torus is made of multiple components that operate at different spatial scales. 
Radially inflowing clouds and starburst regions are occurring at the same scales as the large-scale torus (Hicks et al. 2009), changing its appearance and properties with time. AO-assisted IFUs and ALMA reveal this global structure on scales of $5-20$ pc. 
The gas accretes down to smaller $\sim1$ pc scales, where a dense and turbulent disk is formed. 
KONA will characterize the large-scale properties of the molecular torus in a statistically meaningful sample of nearby AGN, 
and in some cases, it might be possible to obtain a direct image of the molecular torus (as in NGC 1068) and determine its size, morphology, and kinematics.

\subsection{Trends in Nuclear Properties with AGN Properties and Seyfert Type}\label{trends}

The KONA survey is the first to measure the properties of the stars and gas (molecular, ionized, and
highly ionized) at high spatial resolutions in a statistically meaningful sample of powerful bona fide
nearby AGN. This survey therefore has the potential to reveal trends of the measured nuclear properties,
fueling of AGN, and/or outflow properties with the range of AGN properties such as AGN luminosity or Seyfert type (1 versus 2). This could have profound implications for models explaining the
variety of AGN observables. For example, AGN unification models are predicated on the assumption
that Seyfert type 1s are the same population as type 2s, simply viewed such that the line of sight does not
pass through the central dust structure (e.g., Antonucci 1993). Any observed differences in the
distributions or kinematics of the nuclear stars or gas may reveal characteristics of the outer extent of this
obscuring dusty structure (e.g., Schartmann et al. 2014) or may challenge the notion that Seyfert 1s and 2s
are drawn from the same population but rather represent an evolutionary sequence. 
Furthermore, biconical radial outflow is consistent with the torus version of the unified model of AGN, as a bicone of radiation from a collimating torus is a proposition of the model. 
As part of KONA, it will be possible to measure the inclination of the outflow
axis in each galaxy and create a statistical sample of inclination angles of nearby AGN. This will lead to a
direct and robust test of the orientation hypothesis, with potentially dramatic consequences for our
understanding of the unification scheme.

\section{Sample Selection and Characteristics}\label{sample}

The KONA galaxy sample was built driven by the scientific goals described in Section 2. 


The KONA survey must be composed of only bona fide AGN.
This means that the galaxies in the sample must show the four main characteristics of an AGN: (1) optical emission-line ratios consistent with AGN classification in diagnostic diagrams, (2) unresolved hard X-ray emission in the center of the galaxy\footnote{Since Chandra has the best spatial resolution of any X-ray observatory, this usually corresponds to the $2-10$ keV energy band.}, (3) a compact flat-spectrum radio source spatially coincident with the hard X-ray emission, and (4) they must exhibit high-ionization lines. 
Criterion 1 discards AGNs in LINERs, starburst galaxies, and composite systems. 
Criteria 2 and 3 have been used extensively to find AGN in galaxies; however, many AGN exhibiting either hard X-ray or radio compact sources (or both) do not show high-ionization lines. Furthermore, criterion 2 selects against Compton-thick AGN. We found that criterion 4 is the crucial criterion for selecting bona fide AGN, with all Seyfert galaxies exhibiting near-IR high-ionization lines in the literature also fulfilling criteria $1-3$.  
Furthermore, recent spectroscopic studies at optical and near-IR wavelengths show that AGN-driven outflows are present in all active galaxies exhibiting high-ionization lines (Rodr\'iguez-Ardila et al. 2006, Mazzalay et al. 2010, M\"uller-S\'anchez et al. 2011). If these lines are not present, the NLR kinematics is ambiguous, dominated either by rotation or outflow. 

\begin{figure*}
\epsscale{.99}
\plotone{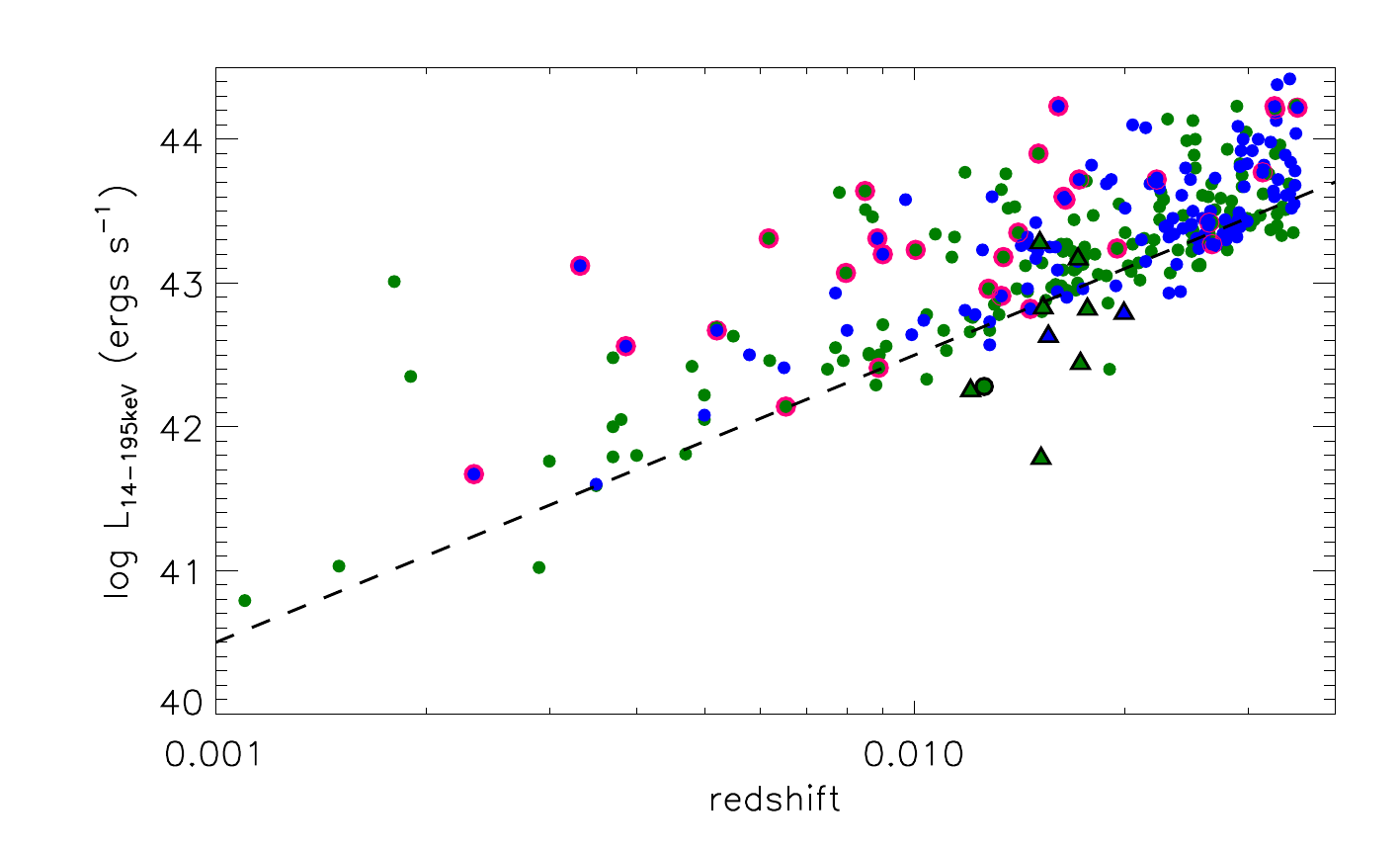}
\caption{Plot of redshift versus $L_{\mathrm{14-195\, keV}}$ for the KONA sample and the 70-month $Swift$-BAT comparison sample (all AGN
in the 70-month $Swift$-BAT catalog with $z<0.035$). Seyfert 1s and 2s are denoted by blue and green, respectively. Those
galaxies in the KONA sample overlapping with the $Swift$-BAT comparison sample are circled in red. The lower limit estimates
of $L_{\mathrm{14-195\, keV}}$ for those KONA galaxies not detected in the 70-month $Swift$-BAT survey, but with existing X-ray observations are represented with black triangles (see text for details). The black circle represents NGC 6967 for which we used the $L_{\mathrm{NIR}}-L_{\mathrm{X}}$ relation from Burtscher et al. (2015). The dotted line is the flux limit for $90\%$ of the sky.}
\label{fig2}
\end{figure*}

Therefore our initial selection consisted of all Seyfert galaxies showing high-ionization lines in existing near-IR spectra (Veilleux et al. 1997, Reunanen et al. 2003, Imanishi \& Wada 2004, Riffel et al. 2006, Rodr\'iguez-Ardila et al. 2011, van der Laan et al. 2013). Then, two observational criteria were imposed: (5) redshift $z< 0.035$ (corresponding to a distance of $\sim150$ Mpc),
and (6) observable from the Keck Observatory ($\delta > -25\degr$). We select objects with $z< 0.035$ because of two reasons: (i) it is the upper limit for [Si VI] 1.96 $\mu$m (the most powerful high-ionization line in the near-IR) to be located in a relatively clean part of the spectrum before the strong atmospheric absorption at $\sim2.0$ $\mu$m, and (ii) it gives a sensible angular scale, so that the extended emission will be well resolved (at $z = 0.015$, the average redshift of the KONA galaxies, $1\arcsec \sim 300$ pc). These three criteria ($4-6$) resulted in 32 galaxies\footnote{These three criteria actually resulted in 36 Seyfert galaxies. However, two of these galaxies were already observed with the instrument SINFONI at the VLT (NGC 1068 and NGC 2992), and the other two show signatures of merger activity (NGC 7674 and HE 1143-1810). Therefore these four galaxies were excluded from the KONA sample. Detailed studies on NGC 1068 and NGC 2992 can be found in M\"uller-S\'anchez et al. (2009), Friedrich et al. (2010) and M\"uller-S\'anchez et al. (2011).} (17 Seyfert 2s and 15 Seyfert 1s). We include an additional two Seyfert 2 galaxies (NGC 3393 and NGC 6967), which met criteria 5 and 6 but had no previous detections of coronal lines in the near-IR (Sosa-Brito et al. 2001). 
These galaxies exhibited other signatures of outflows (e.g., extended [O III] emission; Schmitt et al. 2003) and were found to meet criterion 4 after being observed with OSIRIS (see Section 6). 
We also complement the sample with six more galaxies from the 70-month $Swift$-BAT survey (Baumgartner et al. 2013) that probe the log$L_{\mathrm{2-10\, keV}} > 43.25$ erg s$^{-1}$ regime in our sample\footnote{This regime was chosen because previous studies of the properties of high-ionization lines suggest a relationship between the luminosity of the coronal lines and that of the hard (2-10 keV) X-ray emission (so we expect to detect coronal lines in the more luminous AGN; Rodr\'iguez-Ardila et al. 2011). We found that nineteen galaxies in the $Swift$-BAT survey satisfy our criteria 5 and 6 and have log$L_{\mathrm{2-10\, keV}} > 43.25$ erg s$^{-1}$. We observed the 3 most luminous galaxies of these 19 (Figure 2), and 3 more galaxies that were randomly selected taking into account the inherent complexities of astronomical observations, such as, available time at the telescope, the seeing during the observations, and the  
positions of the galaxies in the sky when the data were taken.} (see Table 1), although they had no prior indication of near-IR high-ionization lines: Ark 120, IC 4329A, Mrk 110, Mrk 590, Mrk 817, and IRAS 05589+2828. In three of these six galaxies are near-IR high-ionization lines detected in our OSIRIS data (see Section 6). The KONA sample of 40 galaxies is presented in Table 1 listed in alphabetical order and separated in two groups (Seyfert 1s and Seyfert 2s)\footnote{In this paper we only classify AGN as Type 1 (broad lines are present in optical spectra) or Type 2 (only narrow lines are present). Intermediate classifications will be considered in forthcoming publications.}.



To place the KONA sample in context, we compare the luminosity distribution to that seen in the 70-month $Swift$-BAT survey (Baumgartner et al. 2013).  This survey measures the $14-195$ keV hard X-ray emission, which is a direct
measure of the AGN emission and widely regarded as the least biased AGN survey. 
Of the 40 KONA galaxies, 30 were detected in the 70-month $Swift$-BAT survey (see Table 1). For these galaxies we estimated the intrinsic $2-10$ keV luminosities using the relation: log$L_{\mathrm{2-10\, keV}} =$ 1.06log$L_{\mathrm{14-195\, keV}} - 3.08$ (Winter et al. 2009). For the rest of the galaxies we adopted the values of $L_{\mathrm{2-10\, keV}}$ from the literature (see Table 1). The only exception is NGC 6967 which has not been observed with X-ray telescopes.




\begin{figure}
\epsscale{1.1}
\plotone{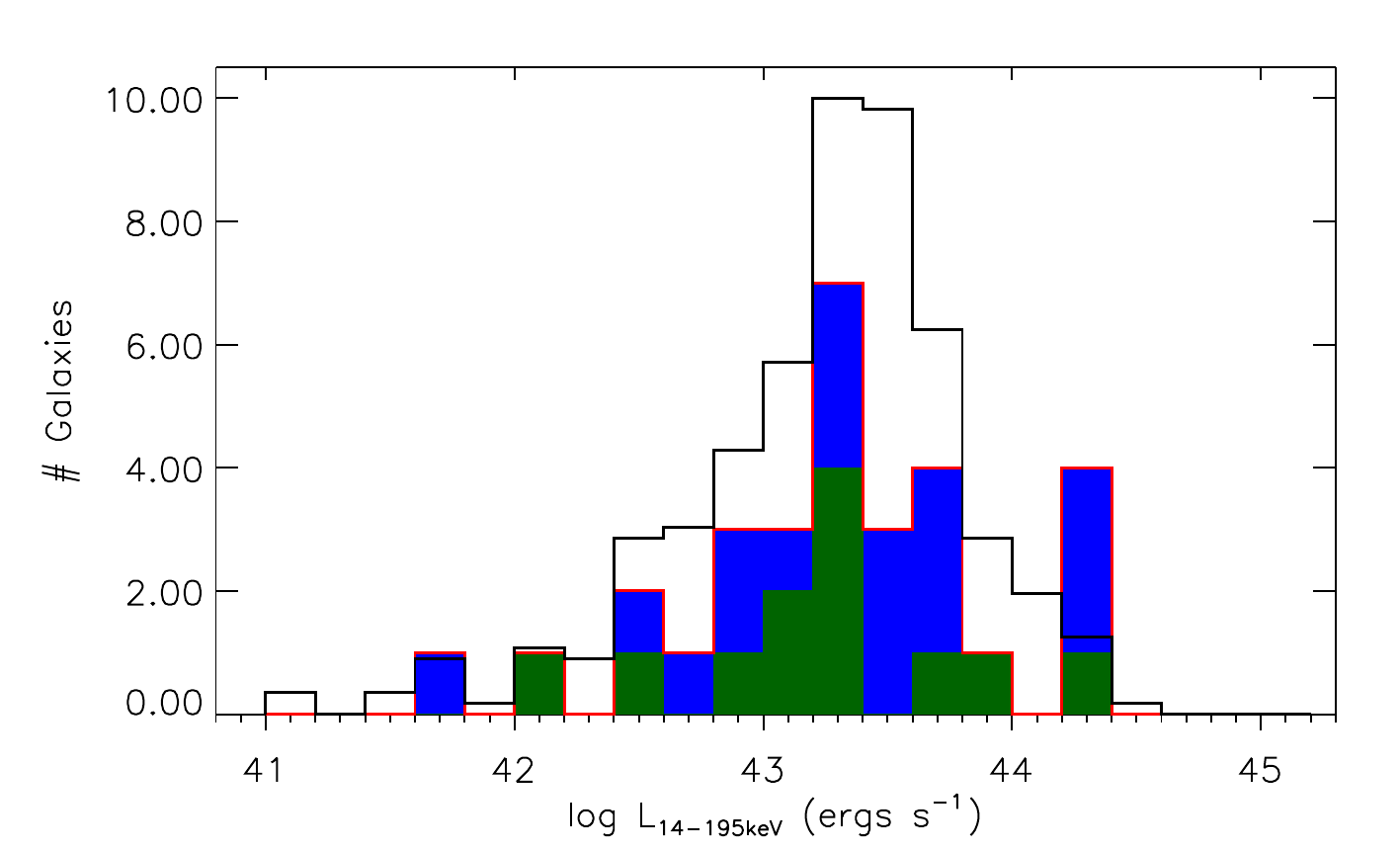}
\caption{Distribution of $L_{\mathrm{14-195\, keV}}$ for the 30 KONA galaxies detected in the BAT 70-month catalogue (see Table 1) compared to the 70-month $Swift$-BAT comparison sample (all AGN in the 70-month $Swift$-BAT catalog with $z<0.035$). 
The galaxies from the KONA sample are shown in color (blue-Seyfert 1s, green-Seyfert 2s) and plotted as the number of galaxies. 
The $Swift$-BAT comparison sample is shown in black and 
is normalized to place the peak at 10. This means that the number of galaxies (292 in total) in each bin is the value shown on the y-axis times the value of 5.6.
}
\label{fig3}
\end{figure}

\begin{figure}
\epsscale{1.1}
\plotone{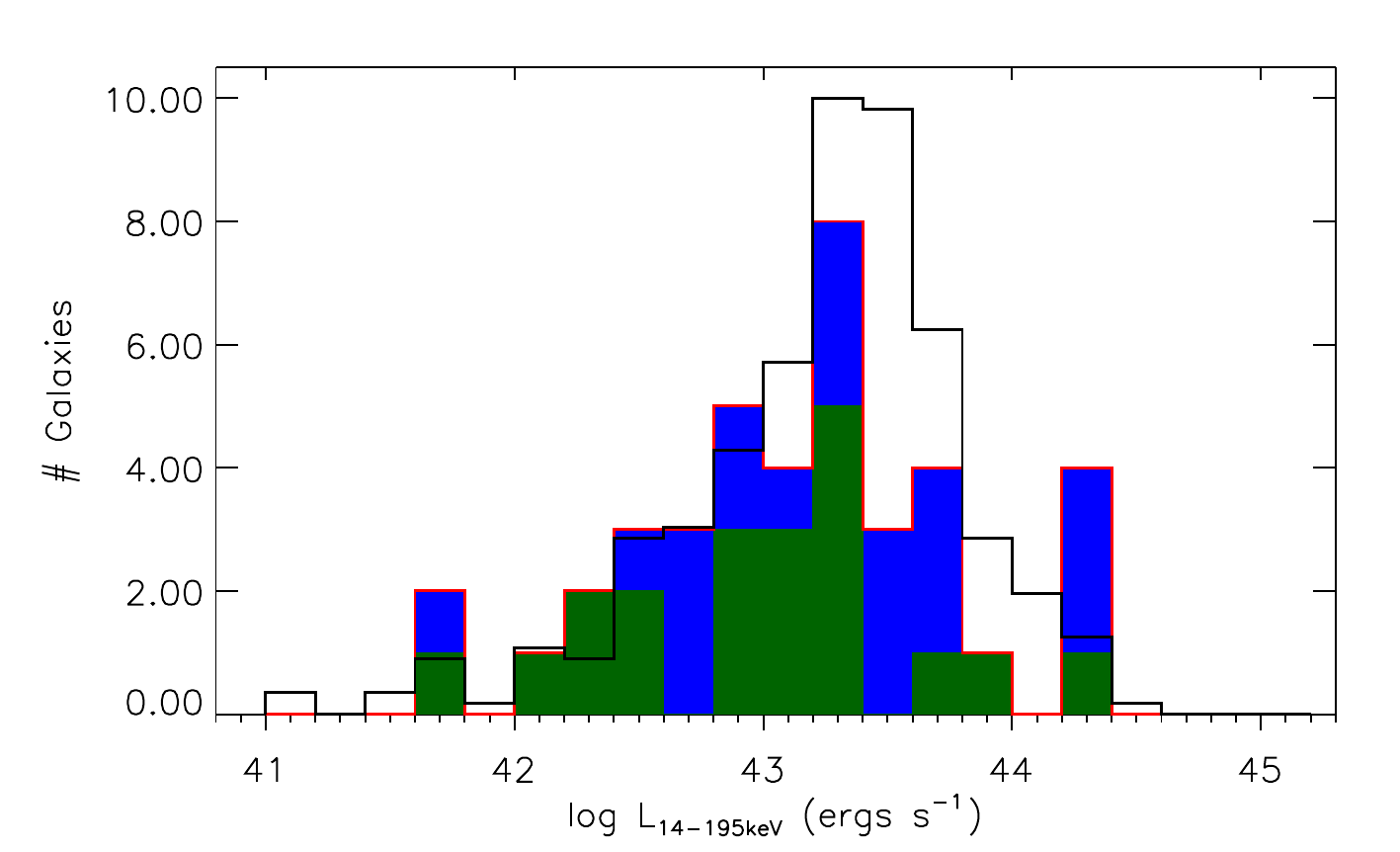}
\caption{Distribution of $L_{\mathrm{14-195\, keV}}$ for the 39 KONA galaxies with measurements of $L_{\mathrm{2-10\, keV}}$  (see Table 1) and NGC 6967 (see text for details) compared to the 70-month $Swift$-BAT comparison sample (all AGN in the 70-month $Swift$-BAT catalog with $z<0.035$). The galaxies from the KONA sample are shown in color (blue-Seyfert 1s, green-Seyfert 2s) and plotted as the number of galaxies. 
The $Swift$-BAT comparison sample is shown in black and 
is normalized to place the peak at 10. This means that the number of galaxies (292 in total) in each bin is the value shown on the y-axis times the value of 5.6.
}
\label{fig4}
\end{figure}

A plot of the distribution of redshift versus $L_{\mathrm{14-195\, keV}}$ is shown in Figure 2 for the full $Swift$-BAT comparison sample with those included in the KONA sample indicated. As shown in Figure 2, those KONA galaxies that overlap with the $Swift$-BAT comparison sample are representative of the $Swift$-BAT sample in both $L_{\mathrm{14-195\, keV}}$ and redshift distribution. The full sample has log($L_{\mathrm{14-195\, keV}}$) ranging from 40.8 to 44.4 erg s$^{-1}$ and mean (median) values of $43.2\pm0.6$ erg s$^{-1}$ (43.3 erg s$^{-1}$). Those galaxies included in
the KONA sample have slightly narrower range in log($L_{\mathrm{14-195\, keV}}$) of 41.7 to 44.2 erg s$^{-1}$ and a consistent mean and median of log($L_{\mathrm{14-195\, keV}}) = 43.3\pm0.6$ erg s$^{-1}$. Including the nine additional galaxies with $L_{\mathrm{14-195\, keV}}$ lower limits based on $L_{\mathrm{2-10\, keV}}$ measurements from the literature (Table 1 and using the Winter et al. relation), and NGC 6967 for which we used the $L_{\mathrm{NIR}}-L_{\mathrm{X}}$ relation from Burtscher et al. (2015), the mean (median) is shifted slightly lower to $43.1\pm0.6$ erg s$^{-1}$ (43.2 erg s$^{-1}$), but still consistent with the mean of the comparison sample, and the range of $L_{\mathrm{14-195\, keV}}$ for the KONA galaxies is unchanged. In terms of the distribution in redshift, the 40 KONA galaxies span a range of $z = 0.0023-0.0352$, with a mean (median) of $z=0.015\pm0.008$. 
The distribution in redshift of the KONA galaxies is slightly lower than those of the $Swift$-BAT comparison sample, but the mean values are consistent, with the $Swift$-BAT comparison sample having a mean (median) of $0.0197\pm0.009$ (0.0202) and range from $z=0.001$ up to the cutoff used to define the comparison sample of $z=0.035$.

For the purposes of interpreting the significance of the comparison of properties for Type 1 and Type 2 AGN, we
compare the properties of the two populations here. As shown in Figures 3 and 4, the Seyfert 2 population has a slightly lower distribution in $L_{\mathrm{14-195\, keV}}$.  For just those KONA galaxies included in the 70-month BAT catalogue the Seyfert 1 and 2 populations have similar distributions, with a mean log($L_{\mathrm{14-195\, keV}}$) for Seyfert 2s of $43.2\pm0.6$ erg s$^{-1}$ compared to the Seyfert 1s mean of $43.3\pm0.7$ erg s$^{-1}$. These distributions are also statistically consistent with those of the full BAT sample for the respective Seyfert 1 and 2 populations.  However, when the estimated $L_{\mathrm{14-195\, keV}}$ values are included, the mean of the Seyfert 2 distribution drops to $43.0\pm1.1$ erg s$^{-1}$ compared to the Seyfert 1 distribution with a mean of $43.2\pm0.7$ erg s$^{-1}$.
In terms of distribution in redshift, the two populations have similar means ($z= 0.017\pm0.010$ and $z=0.014\pm0.006$ for Seyfert 1s and 2s, respectively). 

In summary, Figures 2-4 show that the KONA survey is a representative sample of nearby AGN in terms of hard Xray luminosity and redshift. The sample, which is split between Seyfert 1 and 2, is also unbiased in Seyfert type based on the subsample distributions of these properties. As shown in Figure 2, the sample spans 2.5 orders of magnitude 
in AGN luminosity, with significant luminosity overlap with galaxies at higher redshift. 


\section{Observations and Data Reduction}\label{observations}

\subsection{The OSIRIS Spectrograph}\label{OSIRIS}

AO-assisted IFS of the 40 galaxies in the KONA survey was obtained using the OH Suppressing InfraRed
Imaging Spectrograph (OSIRIS) at the W. M. Keck Observatory. Details on the instrument can be found in Larkin et al. (2006), and
only a brief summary is presented here. OSIRIS is a fully cryogenic integral-field spectrograph, which works with the Keck Adaptive Optics system to reach the diffraction limit of the Keck telescopes. OSIRIS uses an array of infrared transmissive microlenses to sample a rectangular field of view (FoV) at four different spatial scales (0.02, 0.035, 0.05 and 0.1$\arcsec$ pixel$^{-1}$). The microlenses constitute the IFU of the instrument, creating micropupils, which are imaged and dispersed through a classical spectrograph. 
The individual spectra are packed very close together in a Hawaii II HgCdTe
detector (with $2048\times2048$ pixels and 32 output channels). 
OSIRIS offers a variety of near-infrared ($z$, $J$, $H$, and $K$) broad- and narrow-band filters, all with a moderate spectral resolution of $R\sim3800$. 
The size of the FoV depends on the number of angular resolution elements, which differ for each filter. 

\subsection{Details of the Observations}\label{obs}

The KONA galaxies were observed between 2006 April and 2013 November using both the Natural Guide Star (NGS) and Laser Guide Star (LGS) AO systems (van Dam et al. 2004, Wizinowich et al. 2006). In galaxies with bright compact nuclei, we used the AO system in NGS mode. All the galaxies were observed with the Kbb filter (spectral coverage from 1.965 to 2.381 $\mu$m), and the pixel scales were $0.035\arcsec$ pixel$^{-1}$ or $0.05\arcsec$ pixel$^{-1}$, resulting in FoVs of $0.56\arcsec\times2.24\arcsec$ and $0.8\arcsec\times3.2\arcsec$, respectively. The only exception was NGC 5728 which was observed with the $0.1\arcsec$ pixel$^{-1}$ (FoV of $1.6\arcsec\times6.4\arcsec$), due to bad atmospheric conditions during that night. 

The total on-source integration times varied between 20 and 140 minutes, depending on the nuclear $K-$band magnitude of the galaxy as estimated from 2MASS images, as well as limiting factors inherent to astronomical observations (seeing, AO performance, laser collisions, acquisition in the rectangular FoV, time available at the end of the night, etc.). We observed using the classical OSO pattern (exposures of 300 s or 600 s) with a sky offset of 30 arcsec. Due to the rectangular shape of the OSIRIS FoV, each galaxy was observed at a certain position angle (PA, Table 1). 
We measure the PA in the usual way from north ($0\degr$) to east ($90\degr$). 
The rectangular FoV was usually oriented along the photometric major axis of the galaxy. Twenty galaxies were observed at two different PAs. These galaxies exhibit in $HST$ narrow-band images extended [O III] emission in a direction that is not spatially coincident with the major axis of the galaxy (Schmitt et al. 2003; Table 1). 

We observed A0V-A5 stars usually at the beginning, middle, and end of the night for telluric correction and flux calibration. Several stars were observed each night with different air masses for an adequate telluric correction.

\begin{figure*}
\epsscale{.99}
\plotone{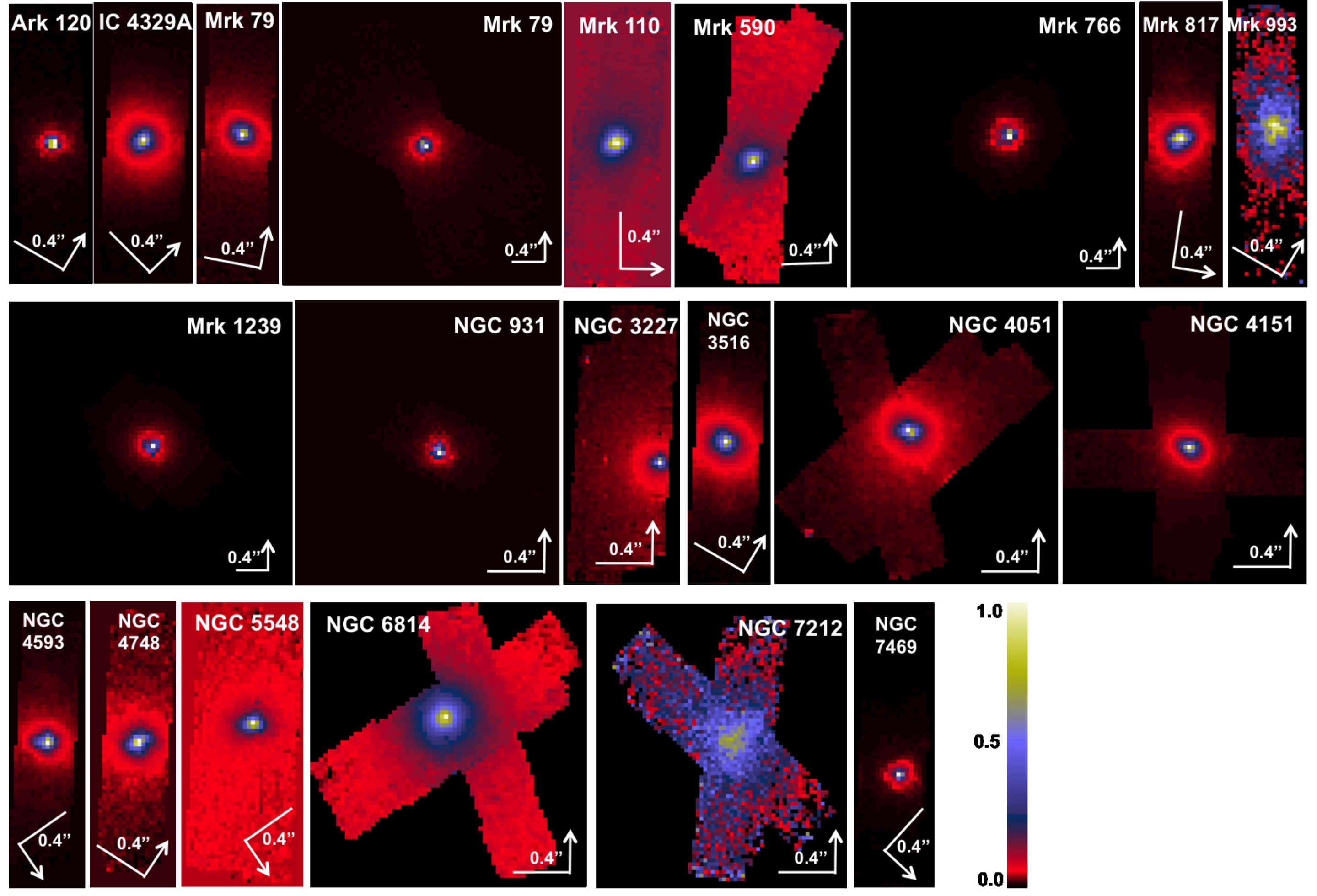}
\caption{Images of the 2.1 $\mu$m continuum emission of the Seyfert 1 galaxies in the KONA sample. The continuum images have been normalized to the peak of emission and are shown in a linear scale. The orientation of the rectangular OSIRIS FoV(s) is given in Table 1 for each galaxy (PA$_1$ and PA$_2$). The arrow indicates the direction north and east is always $90\degr$ counterclockwise from north. The length of the eastern bar is $0.4\arcsec$ in the OSIRIS images.}
\label{fig5}
\end{figure*}

\begin{figure*}
\epsscale{.99}
\plotone{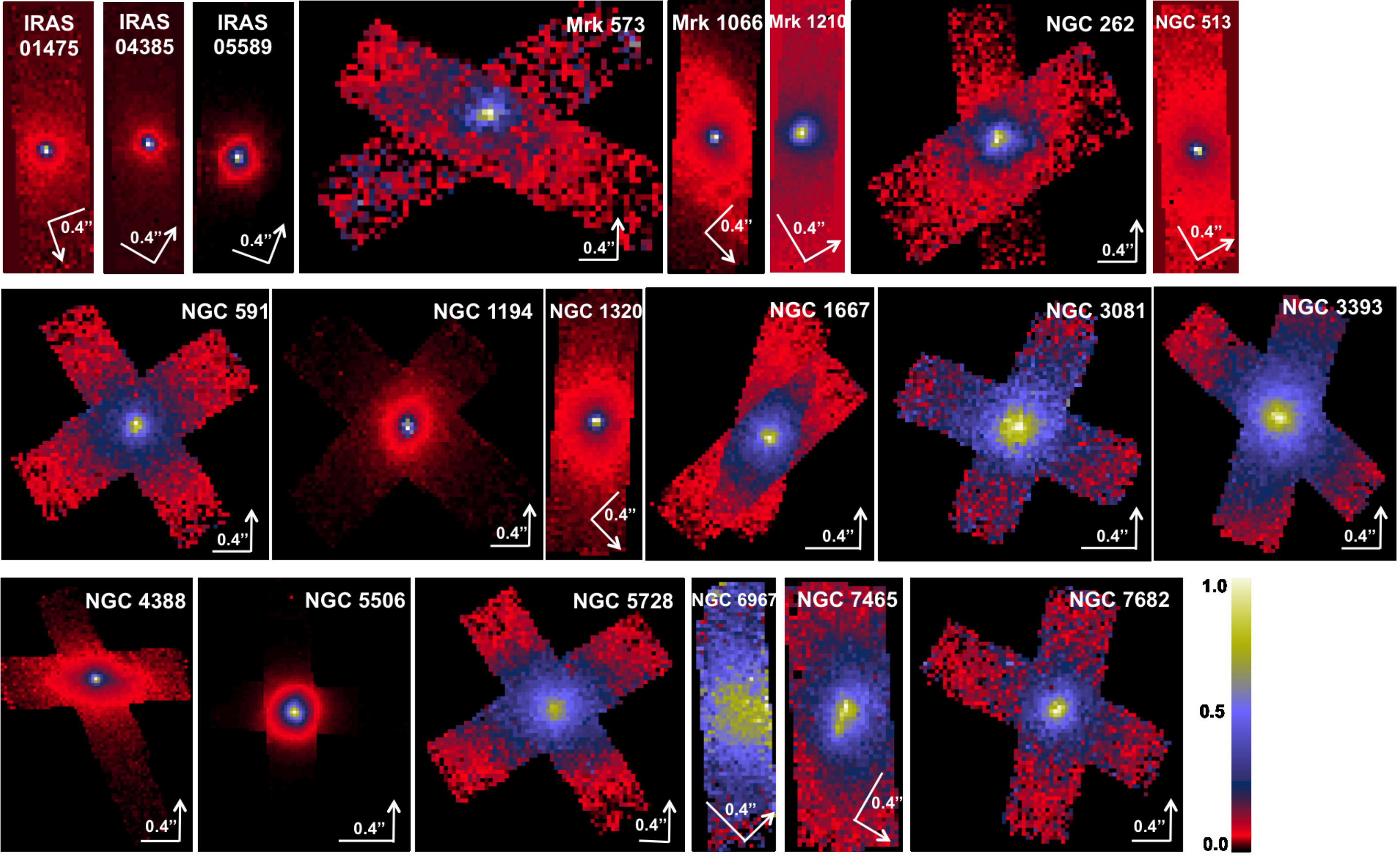}
\caption{Same as Figure 5 but for the Seyfert 2 galaxies.}
\label{fig6}
\end{figure*}

\subsection{Data Reduction}\label{reduction}

The KONA galaxies were reduced using the OSIRIS data reduction pipeline (DRP). The four basic routines of the pipeline: adjust channel levels, remove crosstalk, clean cosmic rays, and correct for dispersion, were always applied to all frames. 
The next step is called rectification and consists of recovering the image at the focal plane (the location of the microlens array) from the intensity image at the detector. 
Rectification matrices were obtained from the Keck observatory webpage for all the different dates of our observations. After rectification, the pipeline then reformatted the 2D spectra into 3D data cubes. 
We subtracted a sky frame from each of the science frames 
using the scaled sky subtraction module, which adjusts the intensity of the OH lines (Davies 2007).

Once the data cubes of the object and the star were constructed, a 1D telluric spectrum was extracted manually from the data cube of the standard star. The spectra of the telluric standard stars were extracted in circular apertures of $0.5\arcsec$. The intrinsic Br$\gamma$ absorption line at 2.166 $\mu$m was removed from the 1D spectrum using a Gaussian function. Then we divided the extracted spectrum by a blackbody spectral distribution with the same temperature of the star. Each individual science data cube was divided by the 1D telluric spectrum and combined using the sigma-clipping average method (``meanclip'') in the mosaicking routine of the DRP. Flux calibration was performed using the magnitude of the standard star as described in M\"uller-S\'anchez et al. (2016). We used the variations of the flux levels between the individual exposures of the same object to estimate the uncertainty in the flux calibration (Erb et al. 2005). 
We find that flux levels between exposures vary by $\sim20\%$. The estimated fluxes were further cross-checked with the magnitudes obtained from 2MASS images in $3\arcsec-5\arcsec$ apertures. We find that the fluxes varied by $15-20\%$ between the different data sources. 

\subsection{PSF Estimation}\label{psf}

The point spread function (PSF) of an AO system depends strongly on the atmospheric conditions at the time of the observation. Therefore the most accurate PSFs are usually extracted from point sources (stars) located inside the FoV. 
However, the small FoVs of AO-assisted IFUs usually do not contain unresolved individual stars. A separate star or field is then observed for PSF calibration. These standard star observations must be close in time and air mass to the science frames in order to remove perturbations caused by the constantly changing atmosphere. 
Unfortunately, for most of the KONA galaxies, a star for PSF calibration was not observed immediately after the science frames. Nevertheless, we used the FWHM of the telluric standard stars as an approximation 
of the PSF of each galaxy (usually a lower limit for the width of the PSF).   

Fortunately, as discussed in Davies et al. (2007), there are a variety of ways to estimate the PSF from the science data itself.  
If a broad component of Br$\gamma$ is present (FWHM $> 1000$ km s$^{-1}$; Davies et al. 2007), this can be used as PSF, since the emission from the BLR is unresolved at our spatial resolution. 
Alternatively, the non-stellar continuum provides a sufficiently good approximation, since at near-IR wavelengths it corresponds to emission of hot dust from the inner part of the torus and is unresolved in our observations (Elitzur et al. 2006). 
Therefore, for Seyfert 1 galaxies, the PSF was estimated from both the broad Br$\gamma$ emission and the non-stellar continuum, and for Seyfert 2 galaxies only the non-stellar continuum was used.

Two-dimensional functions were fit to the data to estimate the PSF. 
For the broad lines, we fitted a two-dimensional Gaussian to all the pixels that contained the broad line emission (normalized to the flux of broad Br$\gamma$ emission). 
For the estimates of the PSF based on the non-stellar emission, we used the code STARFIT (Davies et al. 2007) to obtain an image of the non-stellar continuum. 
Briefly, this method relies on the fact that the intrinsic equivalent width of the $^{12}$CO (2-0) 2.29 $\mu$m bandhead is approximately constant for several starformation histories 
($\sim11-13\AA$; see Davies et al. 2007, Burtscher et al. 2015, and references therein). The observed equivalent width then can be used to determine the dilution of the absorption feature by non-stellar emission. In this case the non-stellar emission is assumed to be attributable to hot dust heated by the AGN, which, at the diffraction limit of the 10 m Keck Observatory, is a point source and thus represents the spatial resolution achieved for the corresponding data cube. Ideally, the intrinsic equivalent width of the $^{12}$CO ($2-0$) bandhead would be determined for each galaxy individually by measuring this value at a radius beyond potential dilution from the AGN continuum (Burtscher et al. 2015).  However, because of the small rectangular FoV of the OSIRIS data, an intrinsic value is difficult to robustly obtain.  Instead we assume an intrinsic equivalent width of $11.1\AA$, which is the median value found by Burtscher et al. (2015) for undiluted sources\footnote{In this paper we use the CO bandheads to obtain an image of the non-stellar continuum and the AGN contribution to the total $K-$band continuum emission ($f_{\mathrm{AGN}}$, see Section 6). The properties of the CO bandheads will be discussed in a forthcoming publication (Hicks et al. 2018, in preparation).}. The resolutions achieved are listed in Table 1. For consistency purposes, the values presented in Table 1 correspond to the PSF estimates based on the non-stellar emission, except for Mrk 110 and Mrk 817, for which we used the spatial extent of broad Br$\gamma$ (in Seyfert 1 galaxies these are consistent with those obtained using the spatial extent of broad Br$\gamma$ emission). In four galaxies (IRAS 05589+2828, Mrk 1239, NGC 5506 and NGC 5548), we could not extract the spatial distribution of the equivalent width of the $^{12}$CO (2-0) bandhead nor the spatial distribution of broad Br$\gamma$. In these cases, we used the FWHM of the $K-$band continuum emission as an upper limit of the achieved resolution (Tables 1 and 2). The errors in the PSF estimates range from $0.01\arcsec$ to $0.04\arcsec$, with a mean value of $0.02\arcsec$ (except in NGC 6967, where the error is large $\sim0.25\arcsec$ due to the absence of a nuclear point source). 

\begin{figure*}
\epsscale{.99}
\plotone{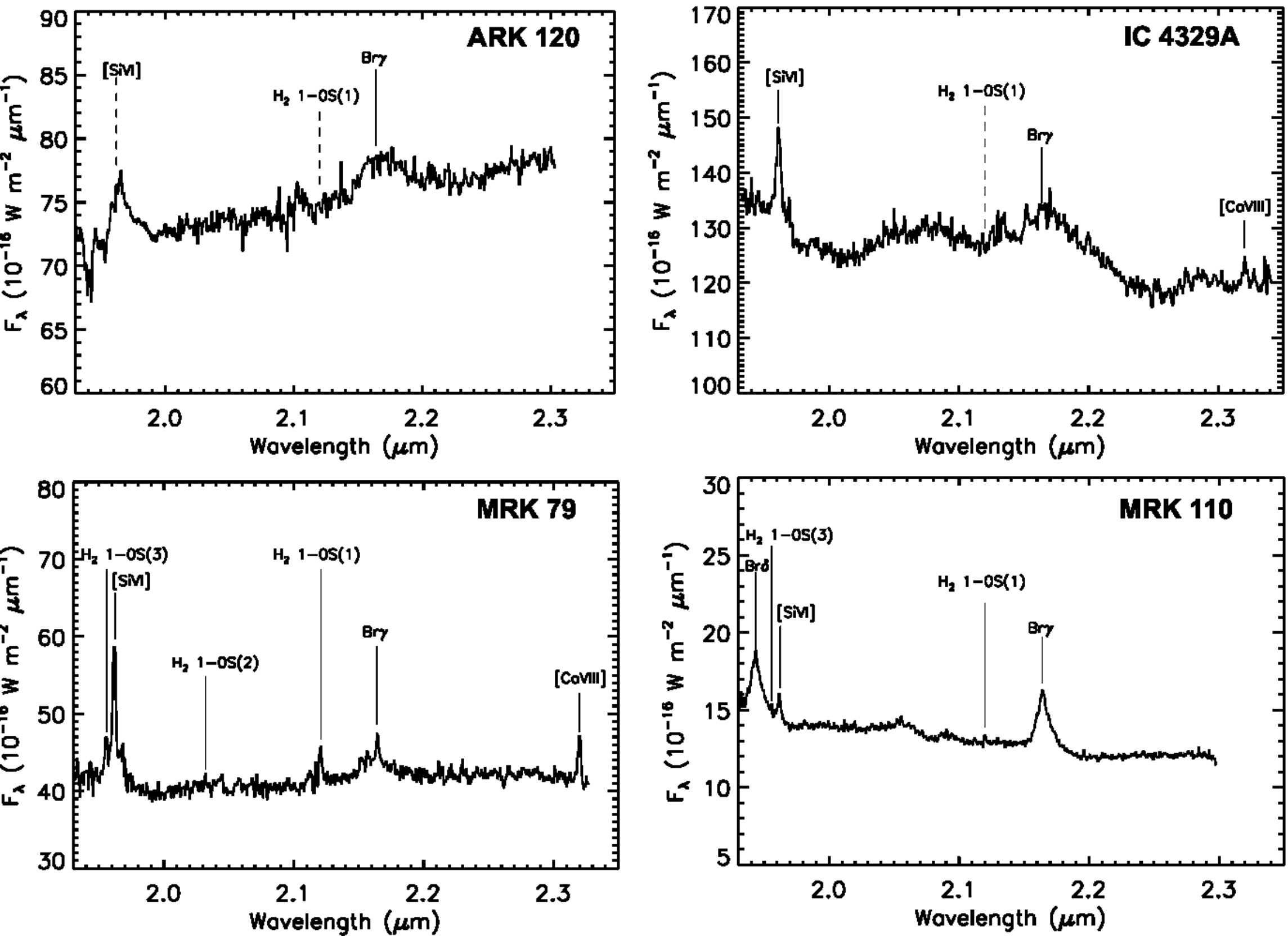}
\caption{$K-$band spectra of the Seyfert 1 galaxies in the KONA sample. The spectrum in each panel was extracted within an aperture of $0.35\arcsec$. Wavelengths are in the rest frame. The data have been smoothed with a median filter of 3 pixels in the spectral direction. All the emission lines detected are indicated with a solid line. Since [Si VI], Br$\gamma$ and H$_2$ 1-0S(1) will be used in future KONA papers to study inflows and outflows in nearby AGN, we indicate their location in the spectrum, even if they are not detected. In these cases they are indicated with a dashed line.}
\label{fig7}
\end{figure*}

\begin{figure*}
\epsscale{.99}
\plotone{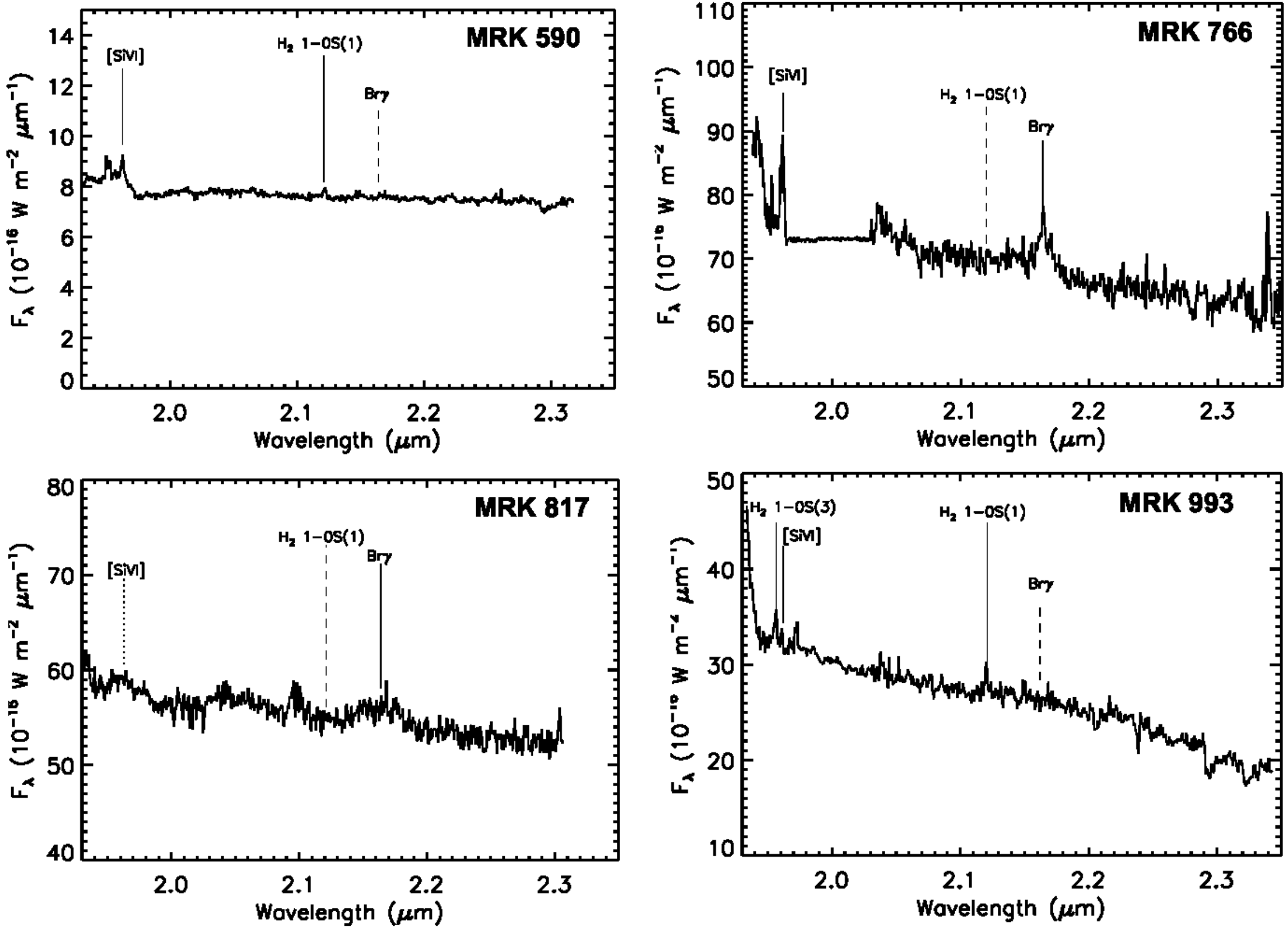}
\caption{Continuation of Fig. 7.}
\label{fig8}
\end{figure*}


\begin{figure*}
\epsscale{.99}
\plotone{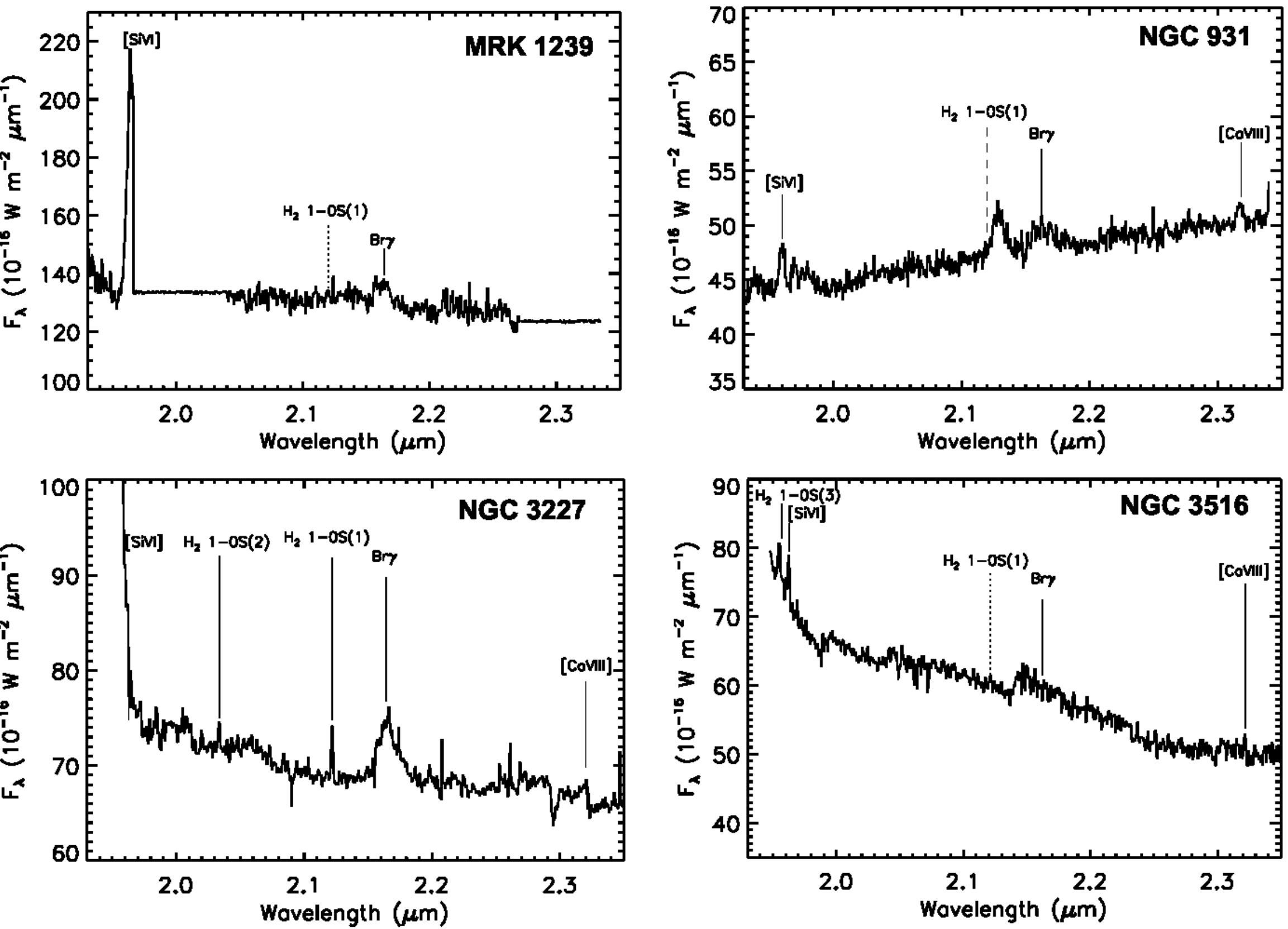}
\caption{Continuation of Fig. 7.}
\label{fig9}
\end{figure*}

\begin{figure*}
\epsscale{.99}
\plotone{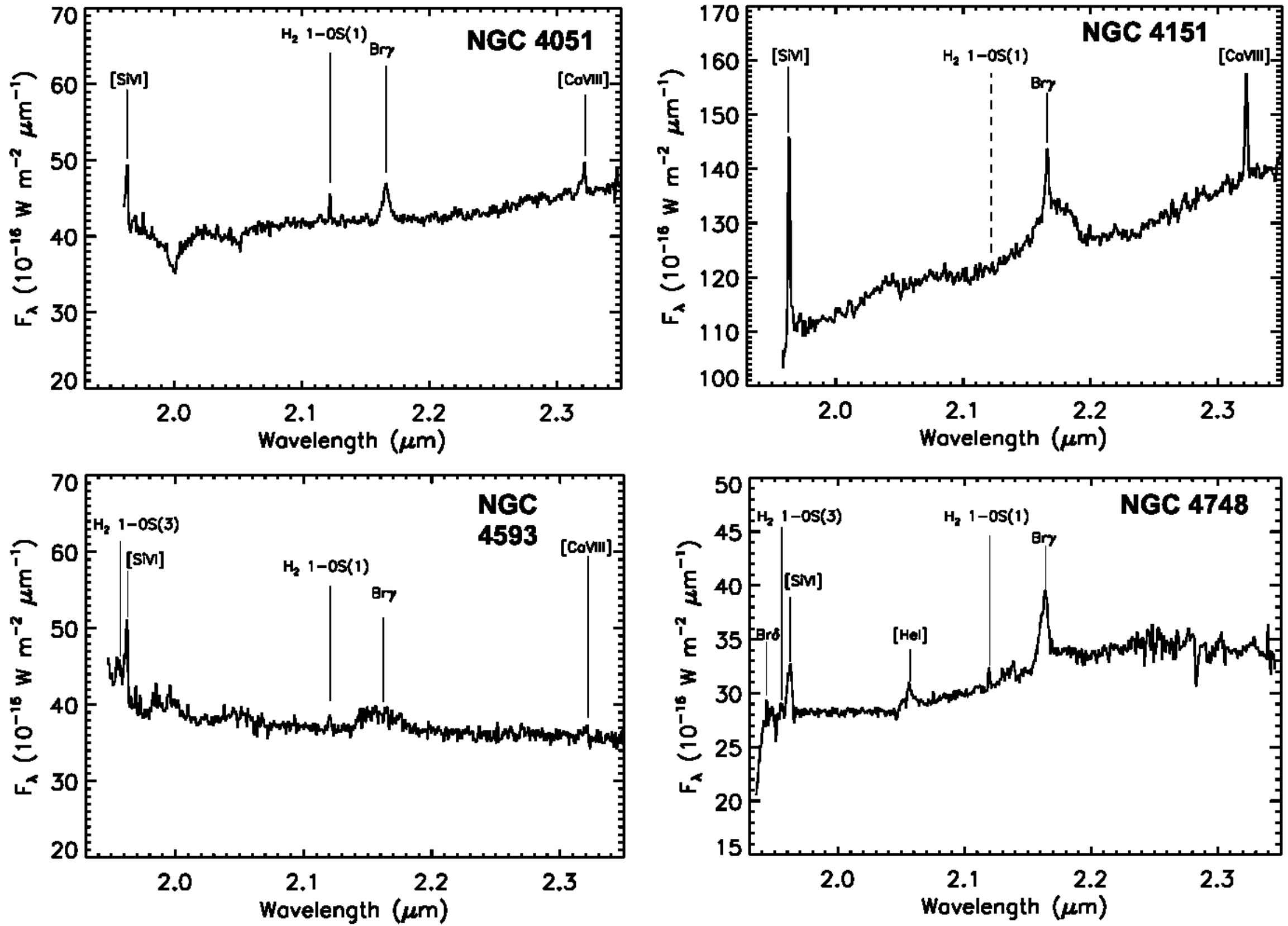}
\caption{Continuation of Fig. 7.}
\label{fig10}
\end{figure*}


\begin{figure*}
\epsscale{.99}
\plotone{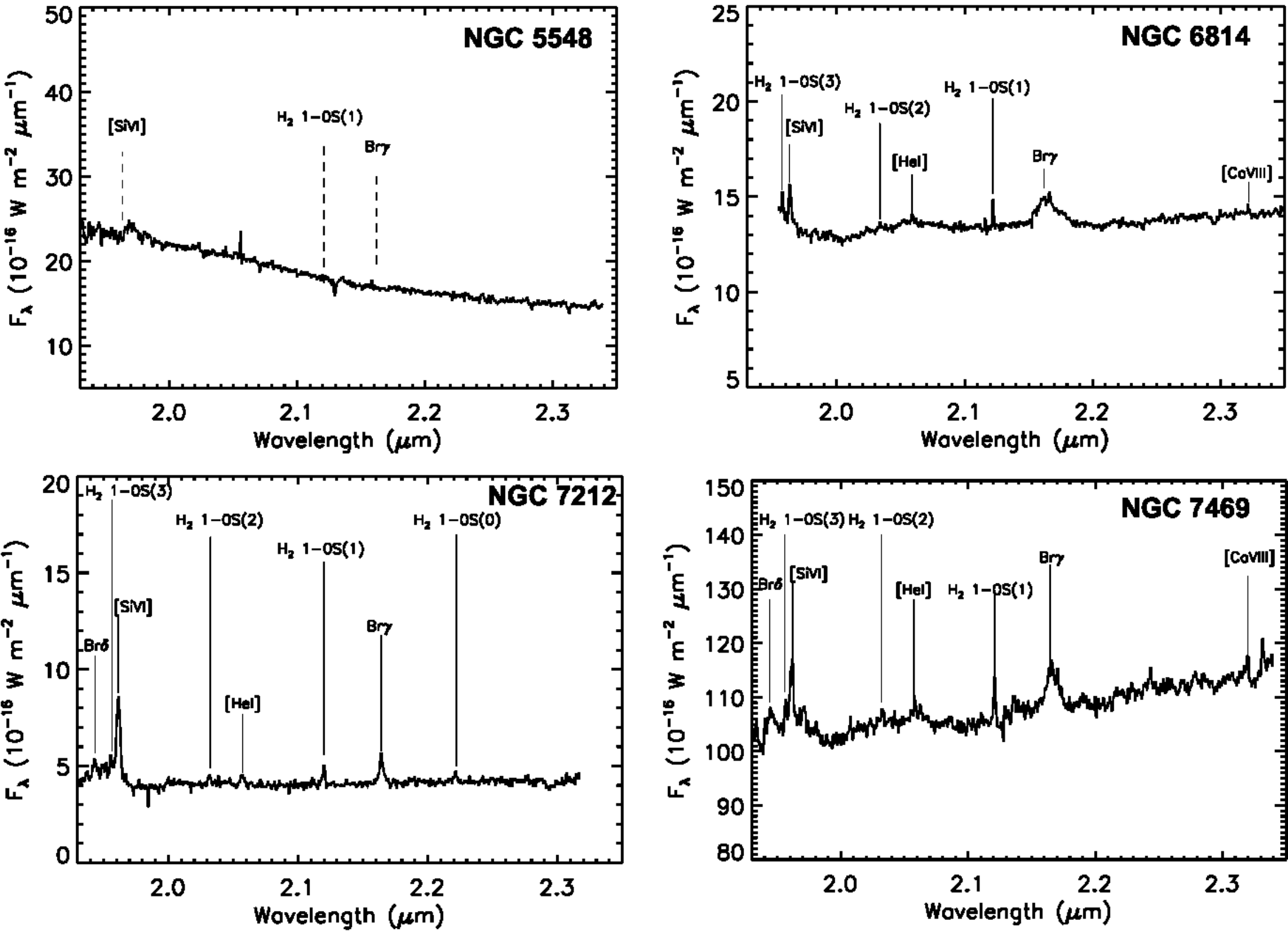}
\caption{Continuation of Fig. 7.}
\label{fig11}
\end{figure*}

\begin{figure*}
\epsscale{.99}
\plotone{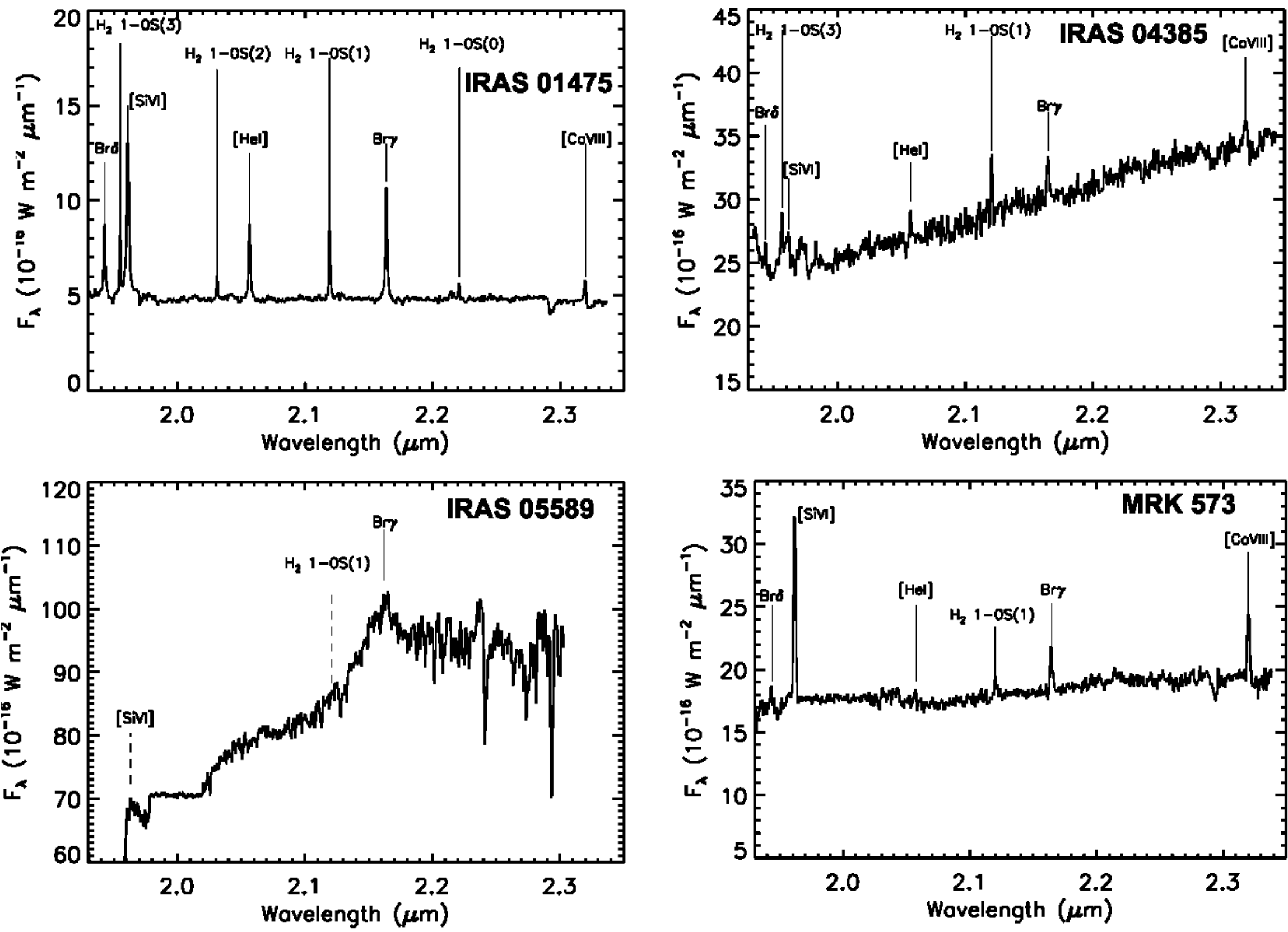}
\caption{Same as Figure 7 but for the Seyfert 2 galaxies}
\label{fig12}
\end{figure*}


\begin{figure*}
\epsscale{.99}
\plotone{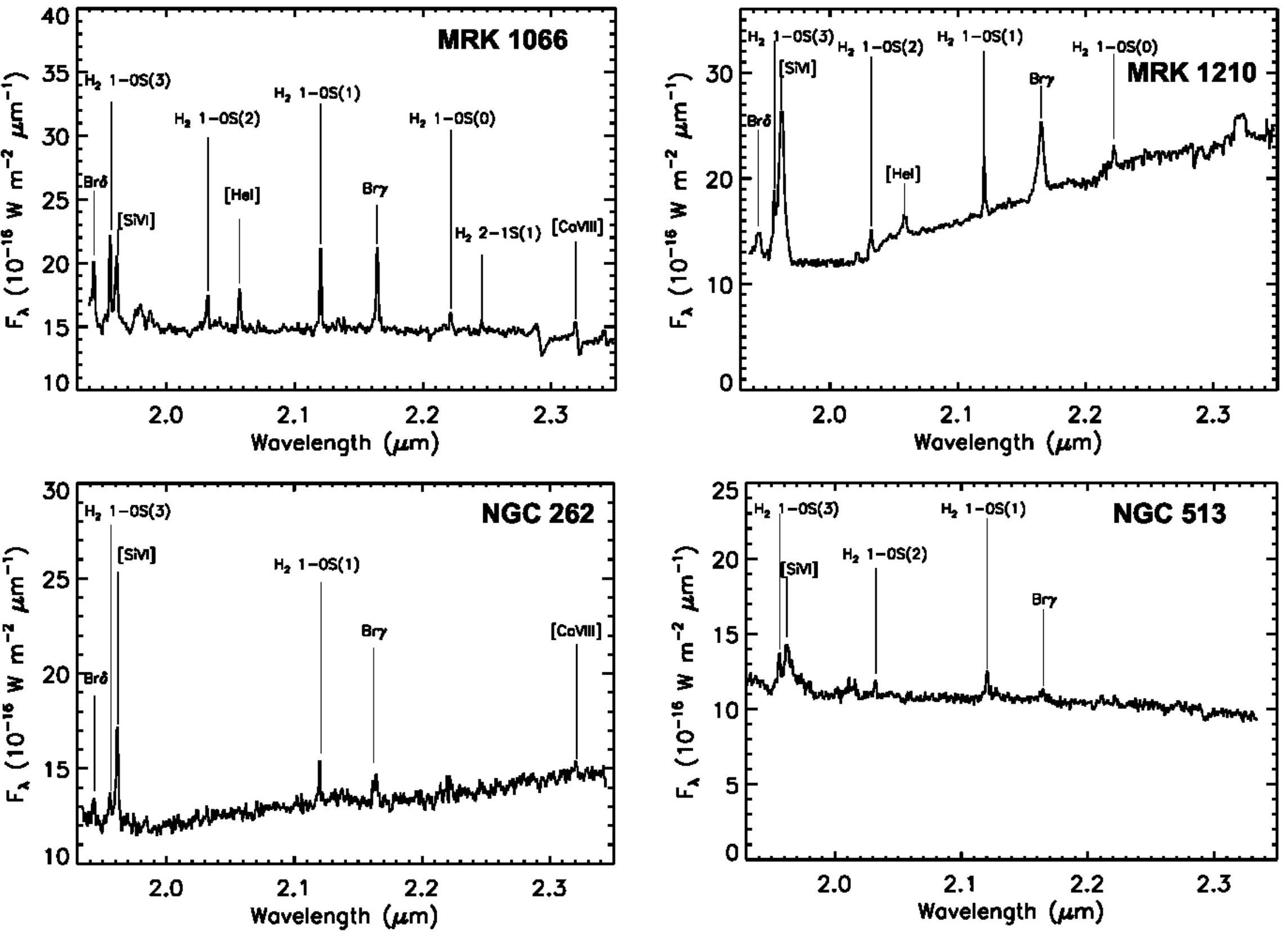}
\caption{Continuation of Fig. 12.}
\label{fig13}
\end{figure*}

\begin{figure*}
\epsscale{.99}
\plotone{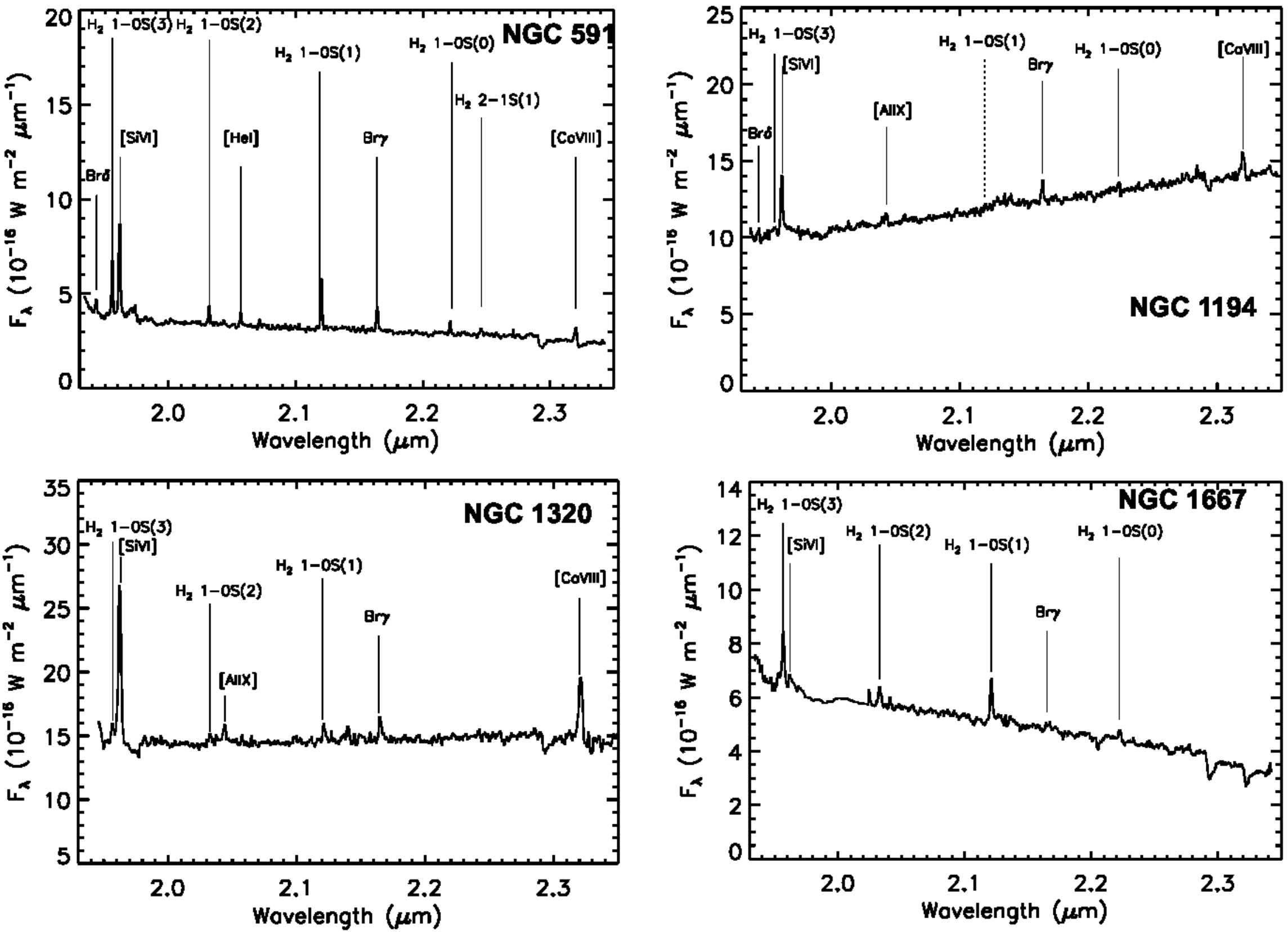}
\caption{Continuation of Fig. 12.}
\label{fig14}
\end{figure*}


\begin{figure*}
\epsscale{.99}
\plotone{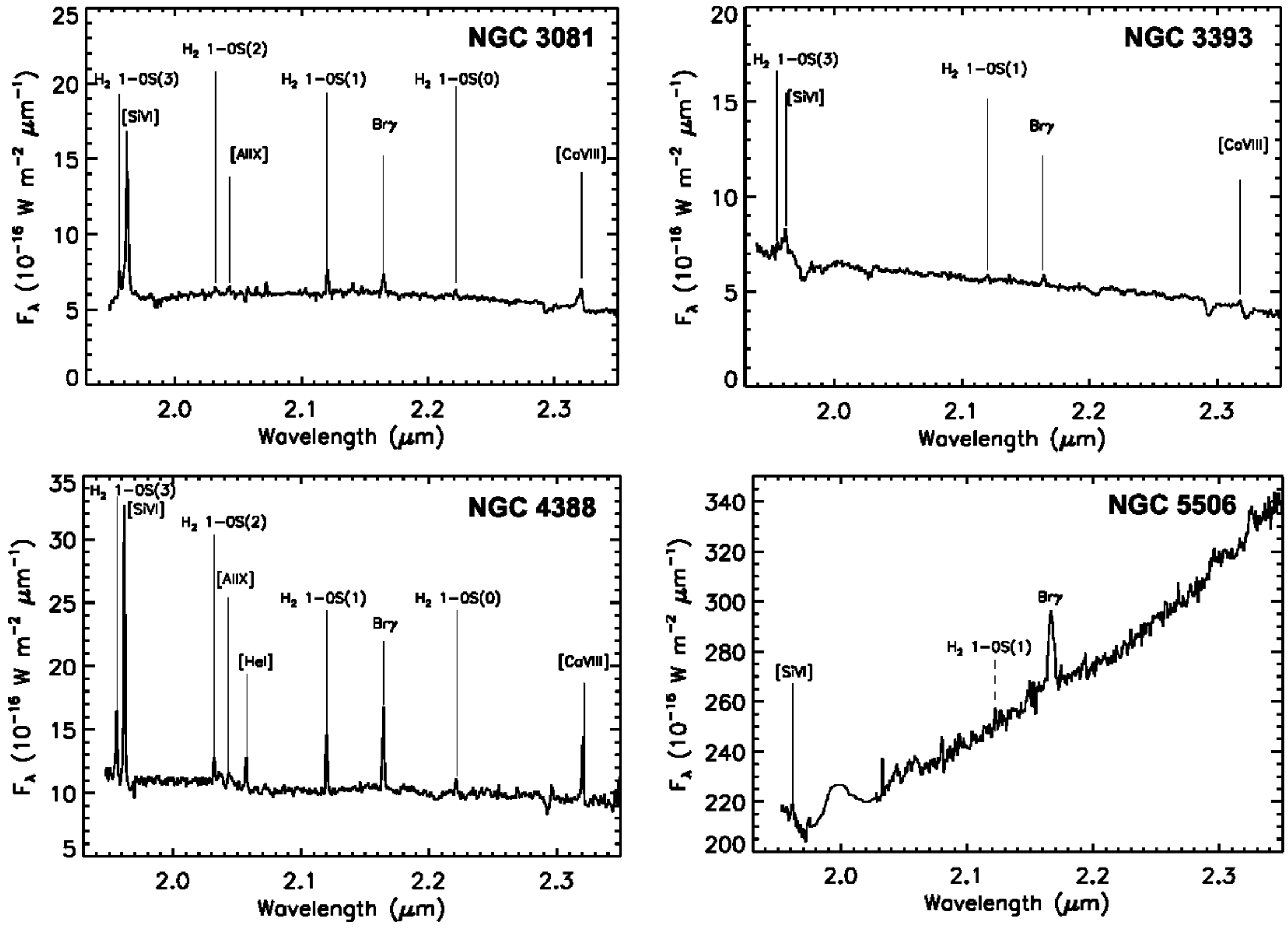}
\caption{Continuation of Fig. 12.}
\label{fig15}
\end{figure*}

\begin{figure*}
\epsscale{.99}
\plotone{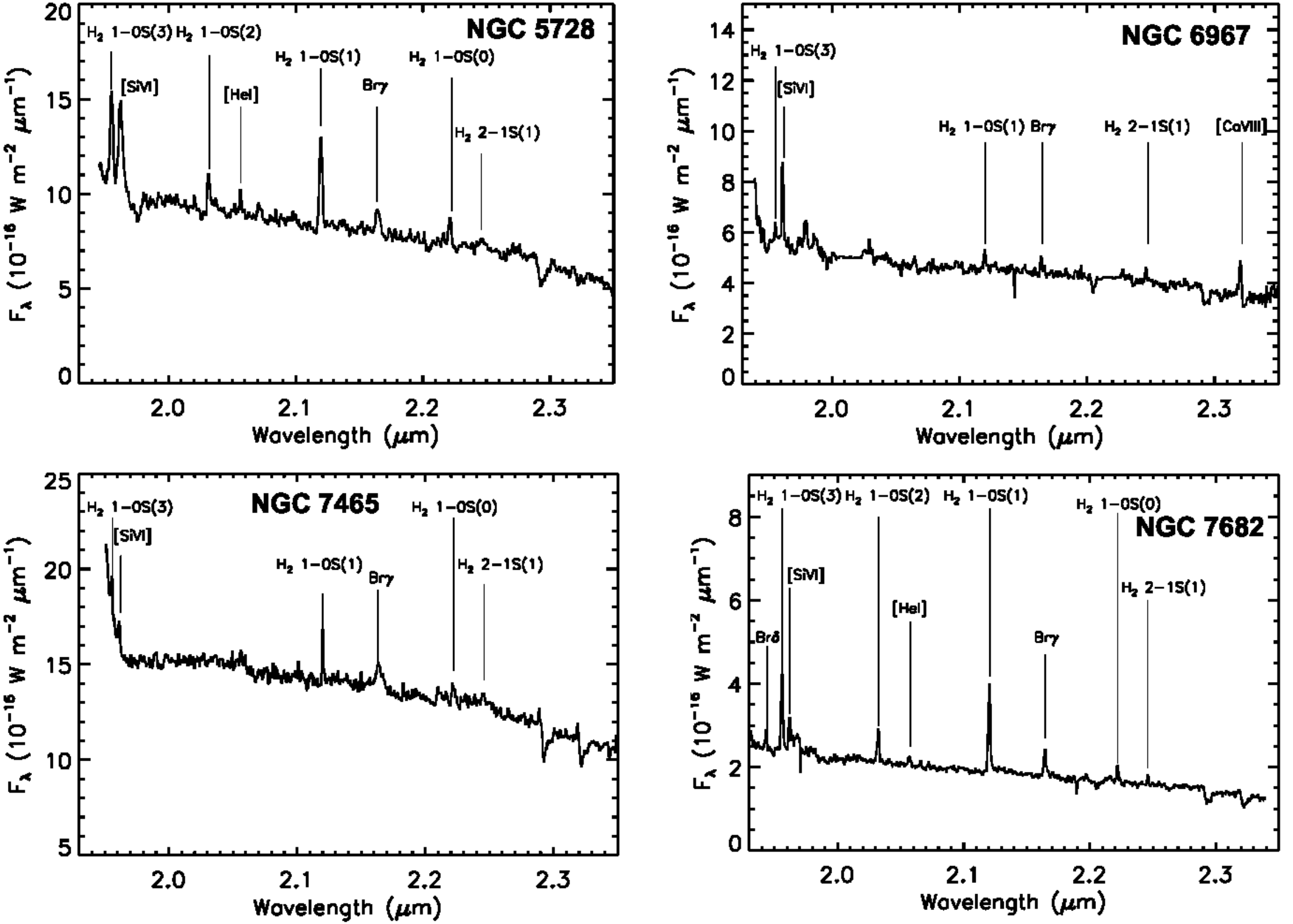}
\caption{Continuation of Fig. 12.}
\label{fig16}
\end{figure*}

\section{Nuclear K-band Properties of the KONA Galaxies}\label{results}

 \subsection{Continuum Emission}\label{continuum}

We present images of the 2.1 $\mu$m continuum emission of the KONA galaxies in Figures 5 and 6. These images show the OSIRIS FoV for each galaxy and the orientation of the data cube(s). All images have been normalized to their respective peak values. The morphologies show a marginally resolved/unresolved source of $K-$band emission in the centers of the galaxies, except in 11 cases: Mrk 993, NGC 262, NGC 591, NGC 1667, NGC 3081, NGC 3393, NGC 5728, NGC 6967, NGC 7212, NGC 7465 and NGC 7682, where the continuum emission is extended. Interestingly, among these 11 galaxies, there are two Seyfert 1s (Mrk 993 and NGC 7212), and NGC 7465 shows signatures of a double nucleus.  
We measured the FWHM and the flux density (magnitude) of the nuclear $K-$band continuum emission, and the results are shown in Table 2. We fitted a symmetric two-dimensional Gaussian to each continuum image and measured its FWHM (FWHM$_{cont}$ in Table 2). For the nuclear magnitudes we used an aperture of $0.35\arcsec$ diameter, which corresponds to 10 pixels for the data cubes observed with the $0.035\arcsec$ pixel scale, and 7 pixels for those observed with the $0.05\arcsec$ pixel scale. Furthermore, this aperture corresponds to $\sim2\times$FWHM of the average spatial resolution of the KONA survey ($0.171\arcsec$, Table 1). This small aperture was chosen as a trade-off between spatial resolution and isolating the AGN near-IR continuum from the host galaxy emission. The error in the magnitudes is dominated by the uncertainty in the flux calibration,  
which we estimate to be about $\pm0.25$ mag (or $\sim20\%$ in flux or luminosity).

Figures 7 -- 16 show the nuclear spectra of the KONA galaxies. The spectra were extracted within an aperture of $0.35\arcsec$ centered at the peak of $K-$band continuum emission.
 
In the majority of the galaxies, the continuum emission 
slopes up or flattens toward longer wavelengths (28 galaxies). This behavior is attributable 
to hot dust emission in the $K-$band, often visible in moderate/high-luminosity AGN (e.g., M\"uller-S\'anchez et al. 2006, M\"uller-S\'anchez et al. 2009, Riffel et al. 2009, Landt et al. 2011). This strongly indicates that we are detecting reprocessed IR emission directly from the AGN torus, with a small contribution from stellar emission. This is predominantly an effect of angular resolution in Seyfert galaxies and demonstrates the advantage of the KONA survey; for instance, previous work has shown that this slope is detected in AO-assisted observations of Mrk 1066 (one of the KONA galaxies), but not in seeing-limited data (Riffel et al. 2010). In 12 galaxies, the continuum emission slopes down toward longer wavelengths. These are 5 Seyfert 1s: Mrk 766, Mrk 993, NGC 3227, NGC 3516 and NGC 5548; and 7 Seyfert 2s: NGC 591, NGC 1667, NGC 3393, NGC 5728, NGC 6967, NGC 7465 and NGC 7682. 
This behavior can be attributed to emission from stars located in the nuclear region. Interestingly, all the Seyfert 2 galaxies with negative spectral slope  and Mrk 993 exhibit extended $K-$band continuum emission, which strongly indicates the presence of nuclear star formation.  

Finally, we estimate the AGN contribution to the $K-$band continuum through a determination of the non-stellar dilution of the $^{12}$CO (2-0) 2.29 $\mu$m bandhead, as described in Section 4.4.  The AGN continuum fraction, $f_{\mathrm{AGN}}$, for each galaxy in which the $^{12}$CO (2-0) bandhead could be measured robustly (31/40 galaxies) is reported in Table 2.  For an additional three galaxies, a lower limit on $f_{\mathrm{AGN}}$ is obtained in a robust measure of the $^{12}$CO (2-0) bandhead in a $1\arcsec$ aperture.  In six galaxies the $^{12}$CO (2-0) bandhead is not detected (in neither the $0.35\arcsec$ or $1\arcsec$ apertures). In these cases we can only assume an upper limit of  $f_{\mathrm{AGN}}$ = 1.0. Interestingly enough, the lowest values of $f_{\mathrm{AGN}}$ are found for the seven Seyfert 2 galaxies with negative spectral slope (NGC 591, NGC 1667, NGC 3393, NGC 5728, NGC 6967, NGC 7465, and NGC 7682), and Mrk 993 exhibits the smallest $f_{\mathrm{AGN}}$ among the Seyfert 1 galaxies (0.48), confirming the presence of significant nuclear star formation in these galaxies.

\subsection{Emission Lines}\label{lines}

We detect several emission lines in the spectra of the KONA galaxies. 
Detailed and quantitative analysis of the 2D properties of the line emission will be presented in separate papers (Hicks et al. 2018, in preparation; Yu et al. 2018, in preparation). Here we present a summary of the emission lines detected in each galaxy and their scientific potential. 
Fluxes of the emission lines are presented in Table 2. In general, 
a flux factor of three above the rms scatter of the spectrum (S/N $>$3) is a criterion for our survey in determining the detection of an emission line. 

The most intense hydrogen recombination line in the 1.95-2.35 $\mu$m range, Br$\gamma$ at 2.165 $\mu$m, is present in the majority of the spectra. Only three galaxies do not show Br$\gamma$ emission (either broad or narrow): Mrk 590, Mrk 993, and NGC 5548 (three Seyfert 1 galaxies). This is not surprising in Mrk 590 and Mrk 993, where the broad H$\alpha$ component is faint, so the even weaker broad Br$\gamma$ emission ($\sim100$ times less flux) is below the detection limit of our observations (see also Riffel et al. 2006). The case of NGC 5548 is intriguing because existing near-IR spectra of this galaxy show broad Br$\gamma$ emission (Riffel et al. 2006). However, the BLR in this galaxy has been reported to be extremely variable (Pei et al. 2017 and references therein), and probably we observed this galaxy during a low-state. Narrow Br$\gamma$ is present in 30 galaxies: all Seyfert 2s and 10 Seyfert 1s (Table 2). In the Seyfert 1 galaxies, the narrow component sits always on top of the broad component. Seven Seyfert 1s do not show a narrow component of Br$\gamma$, and only the broad component is present in the nuclear region (Ark 120, IC 4329A, Mrk 817, Mrk 1239, NGC 931, NGC 3516, and NGC 4593)\footnote{NGC 931 exhibits an asymmetric double-peaked broad Br$\gamma$ profile, which seems to be also present in the long-slit observations of van der Laan et al. (2013).}. 
Interestingly, we detect a broad component of Br$\gamma$ emission in four galaxies that are usually classified as Seyfert 2s: IRAS~05589+2828, Mrk 1210, NGC 5506 and NGC 7465. Particularly, Mrk 1210 and NGC 5506 were previously classified as polarized hidden-broad line region Seyfert 2s (HBLR; Tran et al. 2003). This implies that some HBLR Seyfert 2s are actually obscured Seyfert 1s, in which there exist a definite but relatively modest extinction to the BLR (so that it is hidden at optical wavelengths but not in the near-IR). A detailed comparison between HBLR and non-HBLR Seyfert 2s in KONA will be presented in a forthcoming publication (Yu et al. 2018, in preparation). Br$\delta$ at 1.944 $\mu$m is also detected in some of the galaxies at redshifts $z>0.013$. 

High-ionization transitions ([Si VI] 1.963 $\mu$m, [Al IX] 2.040 $\mu$m and [Ca VIII] 2.322 $\mu$m) are also detected in the majority of Seyferts in the KONA sample. In particular, [Si VI] is detected in 36 galaxies (the only exceptions are Ark 120, IRAS 05589+2828, Mrk 817, and NGC 5548). As discussed in Section 3, we selected for KONA 32 Seyfert galaxies with previous detections of coronal lines in the near-IR. We detected [Si VI] in all of these galaxies except for NGC 5548. This galaxy exhibits a featureless continuum in our OSIRIS data. We report for the first time the detection of [Si VI] in IC 4329A, Mrk 110, Mrk 590, NGC 3393, and NGC 6967. These coronal lines (ionization potential $>100$ eV) are unique tracers of AGN activity and AGN-driven outflows (Mazzalay et al. 2010, M\"uller-S\'anchez et al. 2011, Rodriguez-Ardila et al. 2011). 

Finally, emission from molecular hydrogen (H$_2$) is detected in $85\%$ of the galaxies (34/40). It is important to point out that 
six galaxies without nuclear H$_2$ emission show off-nuclear H$_2$ emission (IC 4329A, Mrk 766, Mrk 817, NGC 3516, NGC 4151 and NGC 5506; Table 2). 
The most powerful H$_2$ emission lines in the near-IR are $1-0$S(3) at 1.956 $\mu$m and $1-0$S(1) at 2.121 $\mu$m, but other transitions of molecular hydrogen are also present in some galaxies. 
The H$_2$ $1-0$S(1) emission line will be used to study the properties of the nuclear molecular gas, its role in obscuring the AGN, its association with the putative torus and the dominant inflow mechanisms in the central tens of parsecs of AGN (Hicks et al. 2018 in preparation).  


\section{Discussion}

As mentioned in Section 3, the most reliable tracer of AGN activity is the hard X-ray emission. Here we compare the total nuclear $K-$band luminosities ($L_K$) and the AGN $K-$band luminosities ($L_K^{\mathrm{AGN}}$) of the KONA galaxies with their intrinsic $2-10$ keV hard X-ray luminosities. In a forthcoming publication, we will investigate possible correlations between the nuclear $K-$band continuum emission and the properties of the several emission lines detected (luminosities and FWHM) with other tracers of nuclear activity such as [O III] 5007 $\AA$ luminosities and 12 $\mu$m emission.

\begin{figure}
\epsscale{1.2}
\plotone{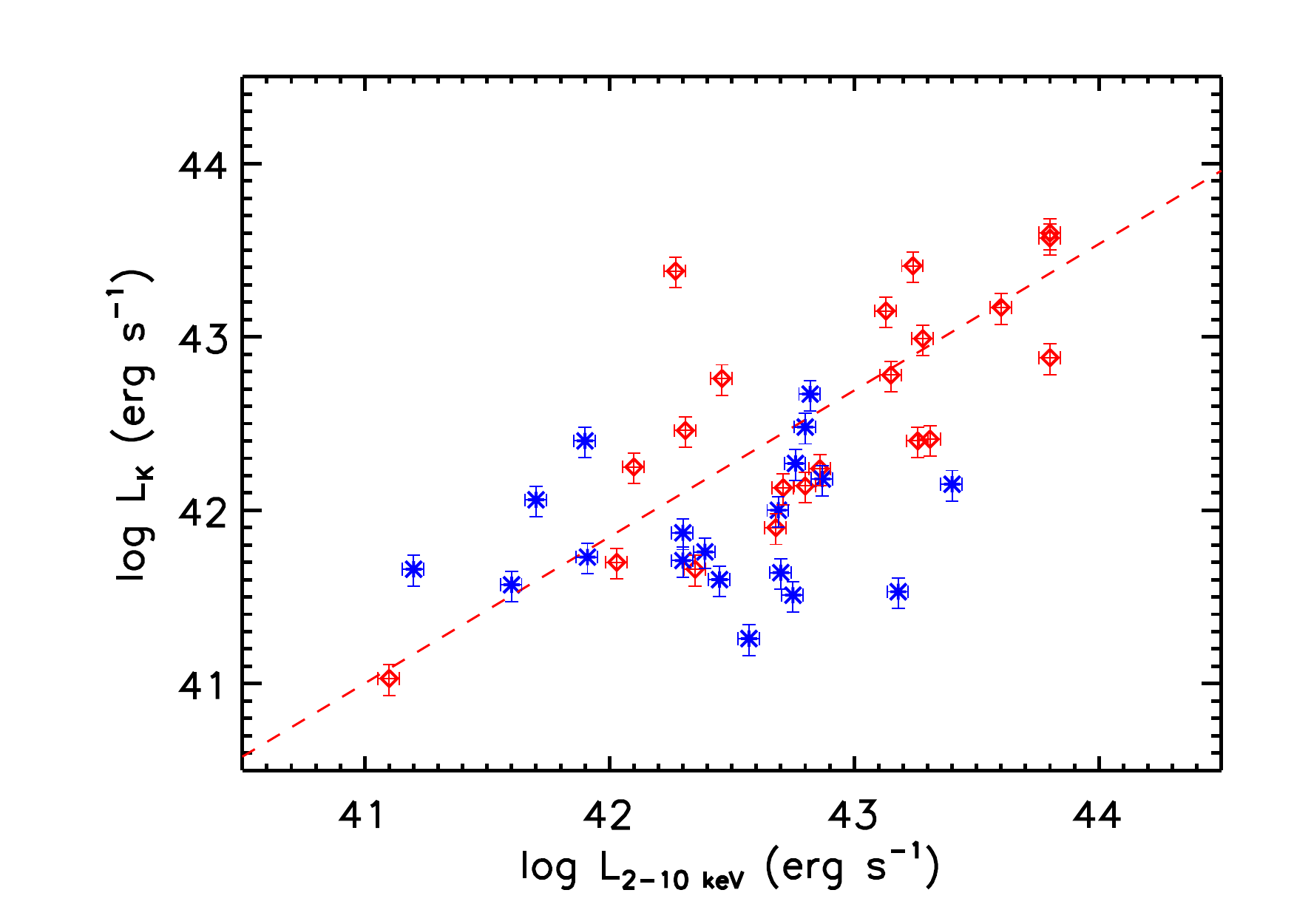}
\caption{Logarithmic plot of total $K-$band luminosity (apertures of $0.35\arcsec$, $L_K$) versus intrinsic $2-10$ keV hard X-ray luminosity for the KONA sample. Red open-squares correspond to Seyfert 1s and blue stars represent Seyfert 2s. A good correlation is found for the Seyfert 1 galaxies. The dashed line represents the best linear fit to the data of the Seyfert 1s (Spearman correlation coefficient of 0.75).}
\label{fig17}
\end{figure}

In Figure 17 we plot the measured total $K-$band luminosities ($L_K$, Table 2) against the intrinsic $2-10$ keV luminosities (Table 1). 
$L_{\mathrm{2-10\, keV}}$ was directly obtained from the $14-195$ keV luminosity in 30 galaxies using the relation: log$L_{\mathrm{2-10\, keV}} =$ 1.06log$L_{\mathrm{14-195\, keV}} - 3.08$ (Winter et al. 2009), and for the galaxies with no measurements of the $14-195$ keV luminosity, we used the values found in the literature (Table 1). Finally, for NGC 6967 we used the $L_{\mathrm{NIR}}-L_{\mathrm{X}}$ relation from Burtscher et al. (2015). 

\begin{figure}
\epsscale{1.2}
\plotone{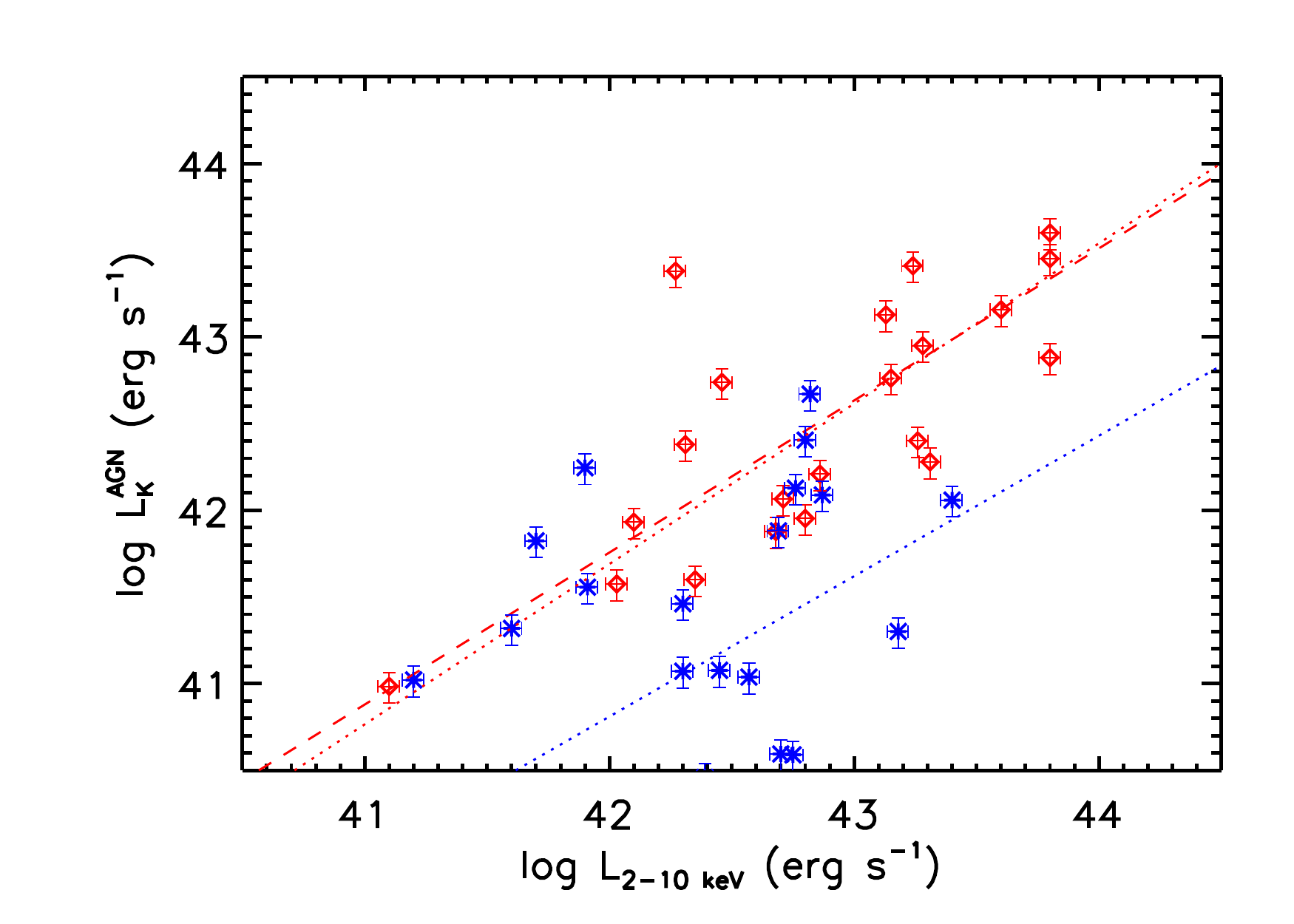}
\caption{Logarithmic plot of AGN $K-$band luminosity ($L_K^{\mathrm{AGN}}$) versus intrinsic $2-10$ keV hard X-ray luminosity for the KONA sample. Red open-squares correspond to Seyfert 1s and blue stars represent Seyfert 2s. A good correlation is found for the Seyfert 1 galaxies. The red dashed line represents the best linear fit to the data of the Seyfert 1s (Spearman correlation coefficient of 0.78). The red and blue dotted lines represent the best fits found by Burtscher et al. (2015) between these two quantities for Seyfert 1s and 2s, respectively. 
}
\label{fig18}
\end{figure}

We computed Spearman's rank correlation coefficients between these two quantities for the full sample of KONA galaxies and for the Seyfert 1 and Seyfert 2 galaxies separately. There is a convincing correlation between hard X-ray and $K-$band luminosity in Seyfert 1 galaxies (Spearman rank correlation coefficient, $R_s$, equal to 0.75). This correlation, and the fact that only two Seyfert 1 galaxies  (Mrk 993 and NGC 7172) do not display unresolved/compact continuum sources, imply that the emission in Seyfert 1s is dominated by non-stellar processes  
on these scales, and that we are measuring direct emission from the AGN. On the other hand, only a weak correlation exists in the Seyfert 2 galaxies ($R_s=0.27$). These galaxies are also less luminous at near-IR wavelengths than Seyfert 1 galaxies of comparable hard X-ray luminosity. 
This lack of a strong correlation suggests either that large extinctions are present toward the Seyfert 2 nuclei, and/or that some fraction of the 2.1 $\mu$m continuum emission is due to nuclear star clusters. 

When we plot the AGN $K-$band luminosities ($L_K^{\mathrm{AGN}}$) against the intrinsic $2-10$ keV luminosities (Figure 18), the correlation found for Seyfert 1s remains practically the same ($R_s=0.78$). This indicates that the AGN $K-$band luminosity in Seyfert 1s can be well approximated by the total $K-$band luminosity on these scales. There is a slightly better correlation between $L_K^{\mathrm{AGN}}$ and $L_{\mathrm{2-10\, keV}}$ for the Seyfert 2 galaxies than 
between $L_K$ and $L_{\mathrm{2-10\, keV}}$, but it is still weak ($R_s=0.45$). For comparison purposes, we also plotted the relations found by Burtscher et al. (2015) between $L_K^{\mathrm{AGN}}$ and $L_{\mathrm{2-10\, keV}}$ for Seyfert 1 and Seyfert 2 galaxies in Figure 18 (red and blue dotted lines, respectively). These authors estimated the $K-$band luminosity of the AGN in apertures of $1\arcsec$ using the same method as the one described in Section 4. As can be seen in Figure 18, the best linear fit to our data of Seyfert 1 galaxies is very consistent with the fit found by Burtscher et al. (2015), despite the differences in the two samples (Burtscher et al. included AGN in LINERs and starburst galaxies), and the fact that Burtscher et al. used apertures of $1\arcsec$ (three times larger than our apertures of $0.35\arcsec$). Based on these results, there seems to be a strong relationship between $L_K^{\mathrm{AGN}}$ and $L_{\mathrm{2-10\, keV}}$ for Seyfert 1 galaxies and has the form: log$L_K^{\mathrm{AGN}}\,=0.925$log$L_{\mathrm{2-10\, keV}}\,+2.84$. If we use the total $K-$band luminosity in $0.35\arcsec$ (or a similar small aperture), the relation is: log$L_K\,=0.9$log$L_{\mathrm{2-10\, keV}}\,+4$. These results demonstrate the importance of high spatial resolution for the study of nearby AGN.

On the other hand, the Seyfert 2 galaxies in KONA appear to be distributed in two groups: one group follows the relation found for Seyfert 1s, and a second group follows the relation found for Seyfert 2s by Burtscher et al. (2015). The fact that the Seyfert 2s are scattered below the relation found for Seyfert 1s, and the apparent division into two groups, 
can be explained by the amount of extinction toward the nuclear region.
As mentioned in Section 5, seven Seyfert 2s show a negative spectral slope and extended $K-$band continuum emission, indicating the presence of nuclear star formation and large amounts of attenuating material (gas and cold dust) toward the nucleus, such that even the hot dust from the AGN torus is strongly obscured (these are the galaxies in group 2). Seven Seyfert 2 galaxies show a flat $K-$band spectrum that indicates moderate amounts of both attenuating material and star formation (these are the galaxies in the transition region between groups 1 and 2). The remaining six Seyfert 2 galaxies (IRAS 04385-0828, IRAS 05589+2828, Mrk 1210, NGC 262, NGC 1194 and NGC 5506) show a positive slope in their nuclear $K-$band spectrum, characteristic of hot dust (M\"uller-S\'anchez et al. 2006, 2009). Since the continuum emission is compact/unresolved in these objects (Figure 6), we interpret these properties as emission from the AGN torus, which is well-correlated with the X-ray luminosity (these are the galaxies in group 1).


These results are consistent with previous studies that have found vigorous star formation in Seyfert galaxies, particularly in Seyfert 2s (Davies et al. 2007, Esquej et al. 2014). Furthermore, the results presented in Figure 17 are consistent with previous spectroscopic studies of nearby AGN, who also found a good correlation between the near-IR and the hard X-ray luminosities in Seyfert 1s, and a weak correlation between these properties in Seyfert 2s (Quillen et al. 2001, Burtscher et al. 2015). It is interesting to note that our data also provide evidence for nuclear star formation in Seyfert 1s as indicated by the galaxies with negative slope and the extended emission in Mrk 993 and NGC 7172 (see Section 5). 

\section{Conclusions}


We have presented the characteristics of the KONA survey, described its science goals and sample selection, 
and used the data to probe the $K-$band nuclear properties of the 40 KONA AGNs, including the detected emission lines, to demonstrate KONA's scientific potential.

With these IFU data of the nuclear regions of 40 Seyfert galaxies the KONA survey will be able to study, for the first time, a number of key topics with meaningful statistics. In this paper we study the nuclear $K-$band properties of nearby AGN. 
We find that the$K-$band ($2.1$ $\mu$m) luminosities of the compact Seyfert 1 nuclei are correlated with the hard X-ray luminosities, implying a non-stellar origin for the majority of the continuum emission. 
We find no strong correlation between $2.1$ $\mu$m luminosity and hard X-ray luminosity for the Seyfert 2 galaxies. The spatial extent and spectral slope of the Seyfert 2 galaxies indicate the presence of nuclear star formation and attenuating material (gas and dust), which in some cases is compact and in some galaxies extended. We detect high-ionization lines in 36 galaxies and for the first time in five galaxies. 
Finally, we find 4/20 galaxies that are usually classified as Type 2 based on their optical spectra exhibit a broad component of Br$\gamma$ emission, and one galaxy (NGC 7465) shows evidence of a double nucleus. 

Our survey also provides a framework for future research on AGN with $JWST$. 
The faintest line flux that we detected at $3\sigma$ in three hours of integration with ground-based AO is $5\times10^{-20}$ W m$^{-2}$. $JWST$ NIRSpec will be able to detect lines as faint as $1\times10^{-21}$ W m$^{-2}$ at $10\sigma$ in a little less than 3 hours. Furthermore, the relationship found between $L_K$ and $L_{\mathrm{2-10\, keV}}$ for Seyfert 1 galaxies would be particularly important for estimating AGN luminosities in objects for which source confusion is difficult in the X-ray regime.


\acknowledgments

F. M.-S. acknowledges financial support from NASA $HST$ Grant HST-AR-13260.001. 
The data presented herein were obtained at the W.M. Keck Observatory, which is operated as a scientific partnership among the California Institute of Technology, the University of California, and the National Aeronautics and Space Administration. The observatory was made possible by the generous financial support of the W.M. Keck Foundation. The authors wish to recognize and acknowledge the very significant cultural role and reverence that the summit of Mauna Kea has always had within the indigenous Hawaiian community. We are most fortunate to have the opportunity to conduct observations from this mountain. The authors thank Hien Tran, Jim Lyke, and Randy Campbell for their support at the W. M. Keck Observatory. 

Facilities: \facility{Keck (OSIRIS)}

\clearpage

\begin{table}
\caption[The Nearby AGN Sample]{Summary of AGN observed with AO and OSIRIS}
\begin{center}
{
\tiny
\begin{tabular}{l c c c c c c c c c c c}
\hline
\hline \noalign{\smallskip}
Object & Type\tablenotemark{a} & $z$\tablenotemark{a} & pc/$\arcsec$  & Pixel Scale & PSF\tablenotemark{b} & T$_{\mathrm{int}}$ & PA$_1$ & PA$_2$ & Date & log $L_{\mathrm{2-10keV}}$\tablenotemark{c} & log $L_{\mathrm{14-195keV}}$\tablenotemark{k}\\
 &  &  &  & ($\arcsec$ pixel$^{-1}$) & ($\arcsec$) & (min) & ($\degr$) &  ($\degr$) & & (erg s$^{-1}$) & (erg s$^{-1}$)\\
\hline \noalign{\smallskip}
Ark~120 	& Sy1 & 0.0327 & 656 & 0.035 & 0.089 & 20 	& 30 & -- & Sep 2006 & 43.80 & 44.22 \\
IC~4329A 	& Sy1 & 0.0160 & 321	& 0.035 & 0.120 & 70 & 45 & --	& Jan 2009, Jan 2010 & 43.85 & 44.25 \\
Mrk~0079  	& Sy1 & 0.0221 & 445 & 0.035 & 0.101	& 80 	& 11 & -- & Jan 2009 & 43.25 & 43.73\\
Mrk~0079 & & & &  0.05 & 0.115 & 80 & 0 & 45 & Mar 2013 & \\
Mrk~0110 	& Sy1 & 0.0352 & 708	& 0.035 & 0.124 & 100 	& 90 & -- & Jan 2009 & 43.82 & 44.23 \\
Mrk~0590 	& Sy1 & 0.0263 & 529 & 0.035 & 0.120 & 70 	& 0 & 170 & Sep 2006, Jan 2009 & 43.31 & 43.76\\
Mrk~0766 	& Sy1 & 0.0133 & 266 & 0.05 & 0.093 & 40 	& 90 & 120 & Mar 2013 & 42.46 & 42.91 \\
Mrk~0817 	& Sy1 & 0.0314 & 631 & 0.035 & 0.120 		& 30 	& 95 & -- & Jan 2009 & 43.26 & 43.70\\
Mrk~0993 	& Sy1 & 0.0155 & 311 & 0.035 & 0.245	& 60 	& 30 & -- & Jul 2012 & 42.10$^d$ & 42.62\\ 
Mrk~1239 	& Sy1 & 0.0199 & 400 & 0.05 &  0.100	& 40 	& 0 & 90 & Mar 2013 & 42.27$^e$ & 42.78\\
NGC~0931 	& Sy1 & 0.0164 & 330 & 0.035 & 0.090 		& 40 	& 60 & 160 & Nov 2013 & 43.15 & 43.58 \\
NGC~3227 	& Sy1 & 0.0038 & 77	& 0.035 & 0.088 	& 140 	& 0 & -- & Mar 2006, Jan 2009 & 42.06 & 42.56\\
NGC~3516 	& Sy1 & 0.0088 & 177	& 0.035 & 0.125 	& 80 	& 28 & -- & Jan 2009 & 42.86 & 43.31\\
NGC~4051 	& Sy1 & 0.0023 & 47 & 0.035 & 0.123 	& 90 	& 20 & 135 & Apr 2006, Jan 2009 & 41.12 & 41.67 \\
NGC~4151 	& Sy1 & 0.0033 & 66 & 0.035 & 0.105 & 60		& 0 & 90 & Mar, Apr 2006 & 42.75 & 43.12 \\
NGC~4593 	& Sy1 & 0.0090 & 180	& 0.035 & 0.114 	& 30 & 145 & --	& Jan 2009 & 42.70 & 43.20\\
NGC~4748 	& Sy1 & 0.0146 & 278 & 0.035 & 0.125 & 60 	& 35 & -- & Mar 2011 & 42.31 & 42.82 \\
NGC~5548 	& Sy1 & 0.0171 & 345	& 0.035 & 0.175 	& 60 	& 150 & -- & Mar, Apr 2006 & 43.25 & 43.72\\
NGC~6814 	& Sy1 & 0.0052 & 100 	& 0.035 & 0.170 	& 60 	& 20 & 135 & Apr, Sep 2006 & 42.35 & 42.67\\
NGC~7212 	& Sy1 & 0.0266 & 549 & 0.035 & 0.252 		& 40 	& 45 & 170 & Jul 2012 & 42.80 & 43.27 \\
NGC~7469 	& Sy1 & 0.0163 & 325 & 0.035 & 0.109 		& 70 	& 132 & -- & Sep 2006 & 43.13 & 43.60 \\
IRAS 01475-0740 & Sy2 & 0.0176 & 354 & 0.05 & 0.097 	& 40 	& 160 & -- & Nov 2013 & 42.30$^f$ & 42.81 \\
IRAS 04385-0828 & Sy2 & 0.0151 & 303 & 0.05 & 0.100  & 40 & 30 & -- & Nov 2013 & 42.79$^g$ & 43.18 \\
IRAS 05589+2828 & Sy2 & 0.0328 & 657 & 0.05 & 0.080 & 20 & 20 & -- & Nov 2013 & 43.80 & 44.23 \\
Mrk~0573 	& Sy2 & 0.0172 & 346	& 0.035 & 0.129 	& 60 	& 60 & 120 & Jul 2012 & 41.90$^h$ & 42.42\\
Mrk~1066 	& Sy2 & 0.0120 & 240  & 0.05 & 0.094 	& 40 	& 135 & -- & Mar 2013 & 41.70$^h$ & 42.24\\
Mrk~1210 	& Sy2 & 0.0140 & 281 & 0.035 & 0.150 	& 60 	& 55 & -- & Mar 2011 & 42.86 & 43.35 \\
NGC~0262 	& Sy2 & 0.0150 & 311 & 0.035 & 0.250 	& 60 	& 5 & 120 & Jul 2012 & 43.39 & 43.85\\ 
NGC~0513 	& Sy2 & 0.0194 & 390 & 0.05 &  0.098	& 40 	& 60 & -- & Nov 2013 & 42.76 & 43.24 \\
NGC~0591 	& Sy2 & 0.0151 & 310 & 0.05 & 0.100  & 40 	& 30 & 120 & Nov 2013 & 41.20$^i$ & 41.77 \\
NGC~1194 	& Sy2 & 0.0133 & 268 & 0.05 & 0.104  & 60 	& 45 & 135 & Nov 2013 & 42.69 & 43.18\\
NGC~1320 	& Sy2 & 0.0088 & 178 & 0.05 & 0.118  & 40 	& 135 & -- & Mar 2013 & 41.90 & 42.42\\
NGC~1667 	& Sy2 & 0.0152 & 315  & 0.035 &  0.350	 & 80  & 150 & 170 & Mar 2013 & 42.31$^i$ & 42.81 \\
NGC~3081 	& Sy2 & 0.0079 & 158 & 0.035 & 0.350 		& 90 	& 60 & 150 & Feb 2011 & 42.57 & 43.07\\
NGC~3393 	& Sy2 & 0.0127 & 255 & 0.05 & 0.174 		& 60 	& 30 & 170 & Mar 2013 & 42.40 & 42.96\\
NGC~4388 	& Sy2 & 0.0084 & 163 & 0.05 & 0.120 		& 60 	& 15 & 105 & Mar 2013 & 43.18 & 43.64\\
NGC~5506 	& Sy2 & 0.0061 & 124 & 0.035 & 0.182 	& 80 & 0 & 90 &  Mar 2011, Jul 2012 & 42.82 & 43.31\\
NGC~5728 	& Sy2 & 0.0100 & 200 & 0.1 &  0.660 & 80 	& 30 & 120 & Mar 2013 & 42.75 & 43.23\\
NGC~6967        & Sy2 & 0.0125 & 252 & 0.035 &  0.735		& 40	& 45 & -- & Jul 2012 & 41.80$^j$ & 42.23 \\
NGC~7465        & Sy2 & 0.0065 & 130 & 0.035 & 0.240 		& 30	& 120 & -- &  Jul 2012 & 41.60 & 42.14\\
NGC~7682 	& Sy2 & 0.0171 & 343 & 0.05 &  0.271	& 40 	& 60 & 170 & Nov 2013 & 42.67$^i$ & 43.10 \\
\hline
\hline
\end{tabular}
}
\tablenotetext{a}{Classification and redshift ($z$) are from the NASA/IPAC Extragalactic Database (NED).}
\tablenotetext{b}{PSF estimates based on the FWHM of the non-stellar emission, except in Mrk 110 and Mrk 817, for which we used the spatial extent of broad Br$\gamma$ emission; and IRAS 05589+2828, Mrk 1239, NGC 5506, and NGC 5548, for which we used the FWHM of the total continuum emission at $2.1\mu$m (see text for details). The typical error in the PSF estimates is $0.02\arcsec$. The individual errors are consistent with those shown for FWHM$_{cont}$ in Table 2, but omitted here for compactness.} 
\tablenotetext{c}{Intrinsic (absorption corrected) $2-10$ keV luminosity estimated using the relation: log$L_{\mathrm{2-10\, keV}} =$ 1.06log$L_{\mathrm{14-195\, keV}} - 3.08$ (Winter et al. 2009), except were noted. References for X-ray luminosities include: d: Guainazzi et al. (2005), e: Liu et al. 2014, f: Huang et al. (2011), g: Lutz et al. (2004), h: Hernandez-Garcia et al. (2015), i: Peng et al. (2006). Errors for these luminosities are taken from the respective references. The median error on $L_{\mathrm{2-10\, keV}}$ is 0.1 dex.}
\tablenotetext{j}{Intrinsic $2-10$ keV luminosity of NGC 6967 estimated using the $L_{\mathrm{NIR}}-L_{\mathrm{X}}$ relation from Burtscher et al. (2015).}
\tablenotetext{k}{Luminosity in the $14-195$ keV band from Baumgartner et al. (2013). For those KONA galaxies not detected in the 70-month $Swift$-BAT survey (letters d--j in the previous column), we used the relation: log$L_{\mathrm{2-10\, keV}} =$ 1.06log$L_{\mathrm{14-195\, keV}} - 3.08$ (Winter et al. 2009).}
\end{center}
\label{table1}
\end{table}

\clearpage

\begin{table}
\caption[$K-$band Luminosities] {K-band luminosities and emission-line fluxes within an 0.35$''$ Aperture}
\begin{center}
{
\tiny
\begin{tabular}{l c c c c c c c c c}
\hline
\hline \noalign{\smallskip}
Object & $D$ & FWHM$_{cont}$  &  $m_K$  & log$(L_K)$ & $f_{\mathrm{AGN}}$ & 
log$(L_K^{\mathrm{AGN}})$ & $F_{\mathrm{H_2}}$ & $F_{\mathrm{Br\gamma}}$\tablenotemark{a} & $F_{\mathrm{[Si VI]}}$ \\
 & (Mpc) & ($\arcsec$) & (mag) & (erg s$^{-1}$) & & (erg s$^{-1}$) & ($\times10^{-18}$W m$^{-2}$)  & 
($\times10^{-18}$W m$^{-2}$) & 
($\times10^{-18}$W m$^{-2}$) \\
\hline \noalign{\smallskip}

Ark 120 & 140 & $0.088\pm0.012$  & 11.86 & 43.57 & $0.76^b$ & 43.45 & $<0.34$ & b & $<0.34$ \\
IC~4329A  & 69 & $0.125\pm0.010$ & 11.30 & 43.17 & 0.97 & 43.16 & $<0.56^d$ & b & $5.49\pm0.44$ \\
Mrk~0079    &  95  & $0.095\pm0.010$ & 12.55  & 42.99  &  0.91  &  42.92 & $0.92\pm0.18$ & $0.97\pm0.16$ & $4.41\pm0.29$ \\
Mrk~0079    &  95  & $0.115\pm0.015$ & 12.54  & 43.00  &  0.91  &  42.93 & $0.92\pm0.18$ & $0.97\pm0.16$ & $4.41\pm0.29$ \\
Mrk~0110   &  151  & $0.144\pm0.010$ & 13.71 & 42.88 & $1.00^c$ & 42.88 & $0.13\pm0.02$ & $0.48\pm0.05$ & $0.44\pm0.03$\\
Mrk~0590   &  113  & $0.124\pm0.023$ & 14.25  &  42.41  &  0.74  &  42.28 & $0.11\pm0.02$ & $<0.05$ & $0.29\pm0.03$  \\
Mrk~0766   &  57  & $0.093\pm0.011$ &   11.93  & 42.76 & 0.95 & 42.74 & $<0.86^d$ & $1.53\pm0.26$ & $4.90\pm0.51$ \\
Mrk~0817   &  135  & $0.120\pm0.021$ & 12.13 & 43.41 &  $1.00^c$  &  43.41 & $<0.33^d$ & b & $<0.33$ \\
Mrk~0993   &  65  & $0.245\pm0.025$  &  12.99  & 42.45  & 0.48 & 42.15 & $0.10\pm0.02$ & $<0.05$ & $0.09\pm0.02$ \\
Mrk~1239   &  81  &  $0.101\pm0.010$  &  11.23  & 43.38 & $1.00^c$ & 43.38 & $<1.20$ & b & $37.30\pm1.30$ \\
NGC~0931   &  71 &  $0.090\pm0.010$  & 12.35   &  42.78 & 0.96 & 42.77 & $<0.18$ & b & $1.75\pm0.23$  \\
NGC~3227   &  17  & $0.089\pm0.010$ & 11.93 & 41.70  &  0.75  & 41.57 & $0.90\pm0.08$ & $2.72\pm0.09$ & $>0.50^e$ \\
NGC~3516   &  35  & $0.126\pm0.010$ & 12.61 & 42.24  &  0.93  &  42.22 & $<0.73^d$ & b & $1.10\pm0.10$  \\
NGC~4051   &  10  & $0.114\pm0.024$  & 12.47 & 41.03  &  0.90  &  41.00 & $0.46\pm0.02$ & $0.42\pm0.02$ & $>1.18^e$ \\
NGC~4151   &  14  & $0.105\pm0.016$  & 11.25 & 41.90  &  0.95  &  41.89  & $<1.25^d$ & $5.6\pm0.36$ & $7.80\pm0.30$ \\
NGC~4593   &  39  & $0.113\pm0.010$  &  12.60 & 42.13  &  $0.86^b$  &  42.04  & $0.50\pm0.08$ & b & $2.20\pm0.20$ \\ 
NGC~4748   &  61  & $0.123\pm0.023$  & 12.83 & 42.46  &  0.83  &  42.36 & $0.29\pm0.04$ & $0.77\pm0.08$ & $2.90\pm0.10$ \\
NGC~5548   &  74  & $0.175\pm0.021$ & 13.29 & 42.40  & $1.00^c$  &  42.40 & $<0.35$ & $<0.35$  & $<0.35$  \\
NGC~6814   &  22  & $0.171\pm0.012$ & 13.71 & 41.23  &  0.87  &  41.17 & $0.20\pm0.02$ & $0.26\pm0.02$ & $0.50\pm0.04$  \\
NGC~7212   &  114  & $0.252\pm0.028$ & 14.98 & 42.14 & 0.65  &  41.97 & $0.22\pm0.03$ & $0.54\pm0.04$ & $1.39\pm0.07$ \\
NGC~7469   &  72 & $0.086\pm0.010$ &  11.46 & 43.15  &  0.95  &  43.14  & $1.40\pm0.22$ & $0.82\pm0.18$ & $3.30\pm0.30$ \\
IRAS~01475-0740   &  76  & $0.122\pm0.016$ & 14.77 & 41.87  &  0.40  &  41.47 & $0.52\pm0.03$ & $1.10\pm0.05$ & $1.75\pm0.05$  \\
IRAS~04385-0828   &  60  &  $0.103\pm0.015$ &  12.88  & 42.48 & 0.84 & 42.40 &  $0.67\pm0.11$ & $0.69\pm0.14$ & $1.09\pm0.21$ \\
IRAS~05589+2828   &  141 & $0.080\pm0.010$   & 11.80   &  43.60 & $1.00^c$ & 43.60 & $<0.92$ & $1.75\pm0.12$ & $<0.92$\\
Mrk~0573   &  74  & $0.131\pm0.035$ & 13.40 & 42.40  &  0.70  &  42.25  & $0.41\pm0.08$ & $0.69\pm0.08$ & $3.18\pm0.09$ \\
Mrk~1066   &  52  & $0.100\pm0.018$ & 13.60 & 42.06  &  0.58  &  41.82 & $1.19\pm0.05$ & $1.35\pm0.06$ & $1.22\pm0.06$  \\
Mrk~1210   &  60  & $0.162\pm0.020$ &  13.41  & 42.18  &  0.81  &  42.09 & $0.75\pm0.05$ & $3.51\pm0.10$ & $6.70\pm0.35$ \\
NGC~0262   &  65  & $0.266\pm0.016$ & 13.72 & 42.15  &  $0.81^b$  &  42.03  & $0.47\pm0.04$ & $0.54\pm0.06$ & $1.21\pm0.08$ \\
NGC~0513   &  83  &  $0.098\pm0.010$ &  13.97 & 42.27  & 0.72 & 42.13 & $0.41\pm0.03$ & $0.33\pm0.05$ & $1.10\pm0.10$ \\
NGC~0591   &  65  & $0.290\pm0.019$ & 14.90 & 41.66  &  0.23 & 41.02 & $0.48\pm0.02$ & $0.25\pm0.01$ & $0.87\pm0.02$ \\
NGC~1194   &  58  & $0.095\pm0.012$  & 13.82 & 42.01  &  0.76  &  41.90 & $<0.12$ & $0.33\pm0.02$ & $0.77\pm0.02$ \\
NGC~1320   &  38  & $0.135\pm0.015$  & 13.61 & 41.73  &  0.67  &  41.55 & $0.30\pm0.05$ & $0.46\pm0.05$ &  $3.55\pm0.08$\\
NGC~1667   &  68  &  $0.350\pm0.025$ &  14.69 &  41.76 & 0.05 & 40.50 & $0.35\pm0.02$ & $0.22\pm0.02$ & $0.27\pm0.03$\\
NGC~3081   &  34  & $0.381\pm0.025$  & 14.50 & 41.26  &  0.64  &  41.06 & $0.33\pm0.03$ & $0.36\pm0.03$ & $1.90\pm0.07$ \\
NGC~3393   &  55  & $0.380\pm0.015$  & 14.70 & 41.60 &  0.33  &  41.11 & $0.12\pm0.02$ & $0.16\pm0.02$ &  $0.36\pm0.03$ \\
NGC~4388   &  36  & $0.139\pm0.013$ & 13.99 &  41.53  &  0.61  &  41.40 & $0.85\pm0.04$ & $1.10\pm0.05$ & $2.86\pm0.11$ \\
NGC~5506   &  27  &  $0.182\pm0.015$  & 10.50  & 42.67 & $1.00^c$ &  42.67 & $<1.80^d$ & $15.20\pm1.0$ & $2.34\pm0.77$ \\
NGC~5728   &  41  & $0.660\pm0.040$ & 14.27 & 41.51  &  0.14  &  40.65 & $1.50\pm0.06$  & $0.52\pm0.06$ & $1.89\pm0.09$ \\
NGC~6967  &  55  & $0.735\pm0.280$ &  14.80 & 41.71  &  0.23  &  41.07 & $0.19\pm0.02$ & $0.15\pm0.02$ & $0.55\pm0.05$ \\
NGC~7465  &  28  & $0.298\pm0.026$  & 13.50 & 41.57  &  0.56  & 41.35  & $0.37\pm0.03$ & $1.10\pm0.11$ &$> 0.17^e$ \\
NGC~7682   &  74  & $0.275\pm0.018$ & 15.44  & 41.64 &  0.09  &  40.64 & $0.40\pm0.02$ & $0.15\pm0.01$ & $0.13\pm0.01$ \\
\hline
\hline
\end{tabular}
}
\tablecomments{Magnitudes, luminosities, and emission-line fluxes integrated in $0.35\arcsec$ apertures. The error in the magnitudes is dominated by the photometric accuracy of our calibration, which we estimate to be about $\pm0.25$ mag (or $\sim20\%$ in luminosity). $L_K$ corresponds to the total $K-$band luminosity in $0.35\arcsec$. $L_K^{\mathrm{AGN}}$ is the AGN luminosity estimated based on dilution of the $^{12}$CO 2.29 $\micron$ absorption feature, assuming an intrinsic CO EQW of 11.1 \AA \, ($L_K^{\mathrm{AGN}}$=$f_{\mathrm{AGN}}*L_K$; see text for details).}
\tablenotetext{a}{Flux of the narrow component of Br$\gamma$. The letter ``b'' indicates that only a broad component of Br$\gamma$ was detected in the spectrum.}
\tablenotetext{b}{S/N too low in $0.35\arcsec$ for a reliable $^{12}$CO 2.29 $\micron$ detection; $f_{\mathrm{AGN}}$ estimated from integrated spectra in $1\arcsec$ apertures.}
\tablenotetext{c}{No $^{12}$CO 2.29 $\micron$ detection (in $0.35\arcsec$ or $1\arcsec$ apertures), upper limit assuming all flux in the $K-$band comes from the AGN.}
\tablenotetext{d}{H$_2$ detected outside the central $0.35\arcsec$.}
\tablenotetext{e}{The profile of the emission line is cut at 1.96 $\mu$m due to the spectral coverage of the instrument. Thus the quoted value is a lower limit.}
\end{center}
\label{table 2}
\end{table}

\end{document}